\documentclass[journal]{IEEEtran}
\IEEEoverridecommandlockouts
\usepackage{comment}
\usepackage[table]{xcolor}% http://ctan.org/pkg/xcolor
\usepackage{mathtools}
\usepackage{bbold}
\usepackage{tabularx}
\usepackage{cite}
\usepackage{float}
\usepackage{multicol}
\usepackage{multirow}
\usepackage{hyperref}
\usepackage{graphicx}
\usepackage{amsfonts}
\usepackage{amsmath}
\usepackage{color}
\usepackage{epstopdf}
\usepackage{caption}
\usepackage{subcaption}
\usepackage{mathtools, nccmath}
\usepackage{multirow}
\usepackage{bbm}
\usepackage{hhline}
\usepackage{marginnote}
\usepackage{tikz}
\usepackage{enumitem}
\usepackage{dblfloatfix}
\usepackage{comment}
\usepackage{eurosym}
\usepackage{makecell}
\usepackage{cite}
\usepackage{amsmath,amssymb,amsfonts}
\usepackage{algorithmic}
\usepackage{graphicx}
\usepackage{textcomp}
\usepackage{xcolor}
\def\BibTeX{{\rm B\kern-.05em{\sc i\kern-.025em b}\kern-.08em
    T\kern-.1667em\lower.7ex\hbox{E}\kern-.125emX}}

\newenvironment{ldescription}[1]
  {\begin{list}{}%
   {\renewcommand\makelabel[1]{##1\hfill}%
   \settowidth\labelwidth{\makelabel{#1}}%
   \setlength\leftmargin{\labelwidth}
   \addtolength\leftmargin{\labelsep}}}
  {\end{list}}

\renewcommand{\theequation}{ \arabic{equation}}
\begin{document}

“This work has been submitted to the IEEE for possible publication. Copyright may be transferred without notice, after which this version may no longer be accessible.”

\title{Electric Vehicle Aggregator as an Automatic Reserves Provider in the European Market Setting}

\author{I. Pavi\'c, \textit{Student Member}, \textit{IEEE}, H. Pand\v{z}i\'c, \textit{Senior Member}, \textit{IEEE} and T. Capuder, \textit{Member}, \textit{IEEE}
\thanks{The authors are with the University of Zagreb Faculty of Electrical Engineering and Computing (e-mails: ivan.pavic@fer.hr; hrvoje.pandzic@fer.hr; tomislav.capuder@fer.hr). %This paper is funded in part by the Croatian Science Foundation under the projects: Active NeIghborhoods energy Markets pArTicipatION -- ANIMATION, Innovative Modelling and Laboratory Tested Solutions for Next Generation of Distribution Networks -- IMAGINE, and Flexibility of Converter-based Microgrids -- FLEXIBASE.}
}}

\maketitle
\begin{abstract}
%Modern power systems are experiencing a paradigm shift where renewable energy sources are replacing fossil-fueled power plants and where electrification of different consumption processes, e.g. transport sector, is in progress. Domination of renewable sources creates a shortage of flexible sources able to provide ancillary services and the power system operators must find new sources flexibility providers. 
Shift of the power system generation from the fossil to the variable renewable sources prompted the system operators to search for new sources of flexibility, that is, new reserve providers. With the introduction of electric vehicles, smart charging emerged as one of the relevant solutions. However, electric vehicle aggregators face the uncertainty of reserve activation on one side and electric vehicle availability on the other. These uncertainty can have a negative effect on both the aggregators' profitability and their users' comfort. %Batteries are energy constrained technology where reserve activation could violate its constraints if not properly managed.
%Electric vehicles could be the solution as vehicles can be controllably charged to provide the desired service. However, their primary goal is to provide mobility using the stored energy. The reserve provision is highly uncertain and it could negatively affect the state-of-energy of an EV and disable it to carry out their primary goal. 

State-of-the art literature mostly neglects the reserve activation or the related uncertainty. Also, they rarely model European markets or use real balancing data. This paper introduces a new method for modeling the reserve activation uncertainty based on actual data for the European power system. Three electric vehicle scheduling models were designed and tested: the deterministic, the stochastic and the robust one. The results demonstrate that the current deterministic approaches inaccurately represent the activation uncertainty and that the proposed models that consider uncertainty, both the stochastic and the robust one, substantially improve the results.   
%have significant affect on electric vehicles and that proposed formulation can solve this issue.
%This paper provides an insight on how the activation of automatic reserves could affect the state-of-energy of the whole fleet and individual EVs taking into account all relevant technical constraints and financial streams. Several models are proposed and validated in order to solve the uncertainty issues arising from the automatic reserve activation. 
\end{abstract}
\vspace{-1mm}
\begin{IEEEkeywords}
Electric Vehicles, Electric Vehicle Aggregator, Frequency Containment Reserve, Frequency Restoration Reserve, Uncertainty
\end{IEEEkeywords}
\vspace{-3mm}
\section{Introduction}
\vspace{-2mm}
Electrification of the transport sector is underway and electric vehicles (EVs) are rapidly increasing their market share \cite{IEA2019}. Large EV fleets can have an adverse effect on the overall power system if inadequately controlled, e.g. increasing the peak power and the balancing needs. The conventional power system operation is already affected by the heavy penetration of renewable energy sources and decommission of fossil power plants. Thus, it is in the need of new flexibility sources that can efficiently balance the system. Smart EV charging seems to be a promising solution due to the EVs' high storage capability \cite{Irena} and availability during the day \cite{JRC-driving}.
%
%could be one of the solutions as EVs are fast ramping units with relatively large storage and power capacity when aggregated \cite{Irena}. Also, most of the day, personal EVs are parked and thus available to be used for power system needs \cite{JRC-driving}. 

Balancing in the European power systems is based on four types of reserves \cite{ENTSOE}: Frequency Containment (FCR), automatic (aFRR) and manual (mFRR) Frequency Restoration, and Replacement Reserve (RR). The FCR and aFRR are automatically activated reserves (activated upon frequency deviation and automatic generation control (AGC) signal, respectively) with fast response and short but more frequent activation events. 
Generally, there are two value streams when it comes to reserves, the capacity reservation and the activation fees. The reservation fee is paid for the availability of a unit to change its operating point, whereas the activation fee is paid for the actual delivery of that change \cite{ACER}.  
Similar to stationary batteries \cite{Figgener}, EVs are technically better suited for automatic reserves \cite{Newtech} and providing them can yield high revenues \cite{PhDbat}. Therefore, the rest of the paper exclusively focuses on FCR and aFRR. 

%The reservation price for FCR and aFRR is commonly higher than for the manual reserves. The EVs are better suited for fast automatic reserves than for manual reserves and in the rest of the paper the focus will be exclusively on automatically activated reserves.  

Three sources of uncertainty can be linked to the EV energy and reserve provision algorithms: EVs' driving behaviour, prices and reserve activation (RA). Uncertainty on the EV behaviour negatively affects the EVs' availability to provide the planned services and the EV batteries' state-of-energy (SOE). This is usually modelled using behavior scenarios \cite{Alipour}, \cite{Shafie}, \cite{Momber}, \cite{Vag} and often includes the second stage re-dispatching measures \cite{Sanchez}, \cite{Bessa}, \cite{Goebel}, \cite{Hasanpor}.
%The most important constraint for the EV reserve provision is its state-of-energy (SOE) status, which is mostly affected by the EV's uncertain behavior. It can be modeled using behavior scenarios \cite{Alipour}, \cite{Shafie}, \cite{Momber}, \cite{Vag} and solved using a two-stage model, usually modeling the day-ahead (DA) and intraday (IDM) markets \cite{Sanchez}.
%The same IDM could be used to avoid DAM plan disturbances caused by reserve activations. 
%The stochastic EV behavior can also be tackled with sequential algorithms from day- to hour-ahead of the delivery where the hour-ahead stage takes new EV information to rearrange its prior charging plans \cite{Bessa}. 
%Interesting approach is to focus directly on the reserve markets, where the intraday market is used to hinder the uncertainty of EV users and to guarantee the reserve delivery \cite{Goebel}. 
The second uncertainty stream are price uncertainties that are commonly addressed through price-taker models using price scenarios \cite{Alipour}, \cite{Momber}, \cite{Vag}, \cite{Shafie} or robust models where prices are determined as a worst-case scenario for the market participant \cite{Liu}, \cite{Kazemi}. Another approach are the price-maker models where the bidding process is designed and the price evolves within the model itself. In such models the participants are paid on the pay-as-bid basis \cite{Goebel}, \cite{aFRR} or by the market clearing price \cite{Energies}. 

The EV behavior and price uncertainties are, in general, well supported by a recent literature. However, there is a gap in the literature in addressing the RA uncertainty. Deterministic modeling of RA \cite{Vag}, \cite{Sanchez}, \cite{Hasanpor}, \cite{Kazemi}, \cite{Sarker} can create problems to EV users as well as to their aggregators \cite{Ivan}. The EV users may suffer from a lower SOE than required for their next trip if the activated up reserve volume was higher than anticipated. An insufficient SOE translates into a decreased comfort level for the EVs users and affects their willingness to participate in reserve provision \cite{Shafie}. On the other hand, if the EV drivers' needs are prioritized, the aggregators may suffer from insufficient energy volumes to back up their day-ahead (DA) plans. %(for arbitrage discharging or up reserve provision). 
Aggregators may experience issues in the opposite direction as well. If down reserve is more frequently activated, the EVs' SOE will be higher than expected and they will not be able to follow their DA schedule. Inability to adhere to the agreed DA schedule causes additional balancing costs \cite{ENTSOE}, whereas the inability to activate the scheduled reserve leads to penalization \cite{Hasanpor}, \cite{Sarker}, \cite{Han} and eventually disqualification from the reserve market participation \cite{Goebel}. The RA uncertainty modeling is thus essential for adequate reserve market participation. One of the possible approaches is stochastic modeling of the RA, as pursued in \cite{Alipour}, \cite{Shafie}, \cite{Vag}, \cite{Han}. However, the randomly generated RA scenarios, instead of the ones based on actual operation data, may not appropriately address the RA uncertainty \cite{Alipour}, \cite{Shafie}, \cite{Han}. On the other hand, employment of publicly available data such as AGC signals (USA-style market \cite{Vag}) or RA data from the European platforms (e.g. ENTSO-E \cite{aFRR}) as scenarios is more appropriate. Following this reasoning, the stochastic model in this paper is designed to utilize such data. 

Robust formulation of uncertainty is another approach rarely used when considering the RA and, to the best of the authors knowledge, only paper \cite{Kazemi} pursues this idea. It considers RA as two values: the number of RA during the day (an integer value) accompanied with a binary value for each call indicating whether the reserve is fully activated or not activated at all. However, such modelling approach is more adequate for manual reserves (mFRR and RR), while automatic reserves require a more rigorous formulation of the uncertainty set. Following this approach, our paper robustly designs the RA uncertainty of the automatic reserves. 

In this paper we model two types of reserves (FCR and aFRR) simultaneously and perform a comparison between their scheduling.  Contrary to the majority of the papers that tackle only the USA-style markets, we focus on modeling the uncertainty of automatic RA for an EV aggregator based on real data stemming from European-style markets. Although European markets were already investigated in \cite{Goebel}, \cite{Hasanpor}, \cite{aFRR}, none of them modeled two types of automatic reserves and none of them based the uncertainty modeling on publicly available RA data. Additionally, this paper for the first time models and compares the RA uncertainty using the stochastic and the robust approach. The contributions of the paper are thus summarized as follows: 
\begin{enumerate}
\item it develops models for aFRR and FCR RA as uncertainty sets (US) based on the information from actual historic reserve data,
\item it integrates a newly created US of aFRR and FCR RA in an EV model and casts it as stochastic and robust linear programs,
\item it simulates a simultaneous provision of aFRR, FCR and DA energy and proves the efficiency and adequacy of the designed models as compared to the deterministic RA model.
\end{enumerate}

The rest of the paper is organized as follows. The mathematical background is formulated and elaborated in Section \ref{sec:math}, case studies and results are presented in Section \ref{sec:case}, while Section \ref{sec:concl} points out the most important findings and concludes the paper.
\vspace{-5mm}
\section{Mathematical Formulation} \label{sec:math}
We design three models to tackle the RA uncertainty: the deterministic, the stochastic and the robust one. After presenting the nomenclature, first we develop the deterministic model, which is later used as a baseline for the stochastic and the robust counterparts.
\vspace{-4mm}
\subsection{Nomenclature}
\subsubsection{Sets and Indices}
\vspace{-1mm}
\begin{ldescription}{$xxxxxx$}
\item [$\mathcal{S}$] Set of scenarios, indexed by $s\in \{1, N_{s}\}$,
\item [$\mathcal{T}$] Set of time steps, indexed by $t \in \{1, N_{t}\}$,
\item [$\mathcal{V}$] Set of vehicles, indexed by $v\in \{1, N_{v}\}$.
\end{ldescription}
\subsubsection{Input Parameters}
\vspace{-1mm}
\begin{ldescription}{$xxxxxxxx$}
\item [$\Delta$] Duration of a time-step [h],
\item [$\Lambda$] Maximum duration of the RA,
\item [$A^{\text{\{UP/DN\}\_\{FCR/aFRR\}}}$] Fixed up/down FCR/aFRR RA ratio,
%\item [$A^{\text{\{UP/DN\}\_\{FCR/aFRR\}}}_{s, t}$] Up/Down FCR/aFRR activation ratio for each scenario $s$,
%\item [$A^{\text{DN\_FCR}}_{s, t}$] Down FCR activation vs reservation ratio for $s$ at $t$,
%\item [$A^{\text{UP\_aFRR}}_{s, t}$] Up aFRR activation vs reservation ratio for $s$ at $t$,
%\item [$A^{\text{DN\_aFRR}}_{s, t}$] Down aFRR activation vs reservation ratio for $s$ at $t$,
\item [$C^{\text{B}}_{v}$] Capital battery cost of EV $v$ [\EUR],
\item [$C^{\text{FCH}}_{v,t}$] Fast charging fee [\EUR/kWh],
\item [$C^{\text{DA}}_{t}$] Day-ahead market electricity price [\EUR/kWh],
\item [$B_{v}$] Battery capacity of EV $v$ [kWh],
\item [$CA^{\text{\{UP/DN\}\_\{FCR/aFRR\}}}_{t}$] Reserve activation fee [\EUR/kWh],
\item [$CR^{\text{\{UP/DN\}\_\{FCR/aFRR\}}}_{t}$] Reserve capacity fee [\EUR/kW/$\Delta$],
% \item [$CR^{\text{DN\_RES}}_{t}$] Reservation fee for either downward FCR or aFRR at $t$ (\EUR/kW),
% \item [$CA^{\text{UP\_RES}}_{t}$] Activation fee for either upward FCR or aFRR at $t$ (\EUR/kWh),
% \item [$CA^{\text{DN\_RES}}_{t}$] Activation fee for either downward FCR or aFRR at $t$ (\EUR/kWh),
\item [$D^{\text{B}}_{\{1/2/3/4\}}$] Battery degradation coefficients,
% \item [$D^{\text{B}}_{1}$] Fixed battery degradation coefficient for higher values of depth-of-discharge.
% \item [$D^{\text{B}}_{2}$] Variable battery degradation coefficient (based on discharged energy) for higher values of depth-of-discharge,
% \item [$D^{\text{B}}_{3}$] Variable battery degradation coefficient (based on depth-of-discharge) for higher values of depth-of-discharge,
% \item [$D^{\text{B}}_{4}$] Variable battery degradation coefficient (based on discharged energy) for lower values of depth-of-discharge,
\item [$E^{\text{RUN}}_{v,t}$] Energy used for driving of EV $v$ [kWh],
\item [$P^{\text{CP\_MAX}}_{v,t}$] Charging point maximum power limit
%for EV $v$
[kW],
\item [$P^{\text{FCH\_MAX}}$] Maximum power limit for fast charging [kW],
\item [$P^{\text{OBC\_MAX}}_{v}$] Maximum OBC power limit for EV $v$ [kW],
\item [$SOE^{\text{\{MIN/MAX\}}}_{v}$] Minimum/maximum SOE of EV $v$ [\%],
\item [$SOE^{\text{T0}}_{v}$] Initial SOE of EV $v$ [\%],
\item [$\eta^{\text{\{RUN/V2G\}}}$] EV driving/V2G discharging efficiency,
\item [$\eta^{\text{\{FCH/SCH\}}}$] EV fast/slow charging efficiency.
\end{ldescription}
\vspace{-2mm}
\subsubsection{Variables}
\begin{ldescription}{$xxxxxx$}
\item [$c^{\text{OTH}}$] Aggregator costs aside from reserve activation [\EUR],
\item [$c^{\text{ACT}}$] Aggregator costs arising from reserve activation [\EUR],
\item [$c^{\text{DEG}}_{v,t}$] Degradation cost attributed to EV $v$ [\EUR],
%\item [$e^{\text{\{BUY/SELL\}\_DA}}_{v,t}$] Energy traded in the DA market for EV $v$ in $t$ [kWh],
\item [$e^{\text{\{BUY/SELL\}\_DA}}_{v,t}$] Energy traded in the DA market [kWh],
%\item [$e^{\text{SELL\_DA}}_{v,t}$] Energy sold from $v$ at $t$ on the day-ahead market (kWh),
\item [$e^{\text{DCH}}_{v,t}$] Energy discharged from EV $v$ [kWh],
\item [$e^{\text{\{FCH/SCH\}}}_{v,t}$] Energy fast/slow-charged to EV $v$ [kWh],
\item [$e^{\text{DEG}}_{v,t}$] Energy used for degradation [kWh],
\item [$e^{\text{OTH}}_{v,t}$] Accumulated en. other than from RA [MWh],
\item [$e^{\text{ACT}}_{v,t}$] Accumulated energy from reserve RA [MWh],
\item [$r^{\text{\{UP/DN\}\_\{FCR/aFRR\}}}_{v,t}$] Reserve capacity sold in reserve markets,% attributed to EV $v$ at $t$ [kW],
%\item [$r^{\text{DN\_RES}}_{v,t}$] Reserved capacity of $v$ at $t$ on either downward FCR or aFRR market,
\item [$soe^{\text{EV}}_{v,t}$] State-of-energy of EV $v$ [kWh].
\end{ldescription}
\vspace{-3mm}
\subsection{Deterministic Model -- DM}
\vspace{-1mm}
\renewcommand{\theequation}{D-\arabic{equation}}
\setcounter{equation}{0}
The RA is usually modeled as an annual average of activations, e.g. \cite{Vag}, \cite{Sanchez}, \cite{Hasanpor}, \cite{Kazemi}, \cite{Sarker}. Similarly, the deterministic model (DM) in this paper, which we use as a baseline, is based on the average annual RA ratio of a particular reserve product. Therefore, the RA ratio the same at all timesteps. The objective function (OF) includes five parts: cost/revenue from energy traded in the DA market, revenue from power sold as FCR/aFRR capacity, V2G battery degradation cost, fast-charging cost and cost/revenue from energy withdrawn/injected as the activated FCR/aFRR. The first four costs other than from RA are assigned to $c^{\text{OTH}}$ defined in eq. \eqref{OF_NRA}, while the costs associated to RA are assigned to $c^{\text{ACT}}$ defined in eq. \eqref{OF_RA}.
Objective function:
\begin{gather} 
\min_{\Xi_{\mathcal{O}}} \quad 
(c^{\text{OTH}}
+
c^{\text{ACT}});
\label{OF}
\end{gather}
is subject to:
\begin{gather} 
c^{\text{OTH}} \! = \! \sum\nolimits_{t=1}^{N_{t}} \! \Big\{ \! \sum\nolimits_{v=1}^{N_{v}} \! \big[
e^{\text{BUY\_DA}}_{v,t} \! \cdot \! C^{\text{DA}}_{t} \!
- \! e^{\text{SELL\_DA}}_{v,t} \! \cdot \! C^{\text{DA}}_{t} \!
\nonumber \\
- r^{\text{UP\_FCR}}_{v,t} \cdot CR^{\text{UP\_FCR}}_{t}
- r^{\text{DN\_FCR}}_{v,t} \cdot CR^{\text{DN\_FCR}}_{t}
\nonumber \\ 
- r^{\text{UP\_aFRR}}_{v,t} \cdot CR^{\text{UP\_aFRR}}_{t}
- r^{\text{DN\_aFRR}}_{v,t} \cdot CR^{\text{DN\_aFRR}}_{t}
\nonumber \\
+ c^{\text{DEG}}_{v,t}
+ e^{\text{FCH}}_{v,t} \cdot C^{\text{FCH}}_{v,t}
\big] \Big\}; 
\label{OF_NRA} \\
c^{\text{ACT}} \! = \! \sum\nolimits_{t=1}^{N_{t}} \! \Big\{ \! \sum\nolimits_{v=1}^{N_{v}} \! \big[
\! - \! r^{\text{UP\_FCR}}_{v,t} \! \cdot \! A^{\text{UP\_FCR}} \! \cdot \! \Delta \! \cdot \! CA^{\text{UP\_FCR}}_{t} \nonumber 
\\
+r^{\text{DN\_FCR}}_{v,t} \cdot A^{\text{DN\_FCR}} \cdot \Delta \cdot CA^{\text{DN\_FCR}}_{t} \nonumber 
\\
- r^{\text{UP\_aFRR}}_{v,t} \cdot A^{\text{UP\_aFRR}} \cdot \Delta \cdot CA^{\text{UP\_aFRR}}_{t} 
\nonumber 
\\
+ r^{\text{DN\_aFRR}}_{v,t} \cdot A^{\text{DN\_aFRR}} \cdot \Delta \cdot CA^{\text{DN\_aFRR}}_{t}
\big] \Big\};
\label{OF_RA}
\end{gather}
\begin{gather}
e^{\text{BUY\_DA}}_{v,t}, e^{\text{SELL\_DA}}_{v,t}, r^{\text{UP\_FCR}}_{v,t}, 
\nonumber  \\
r^{\text{DN\_FCR}}_{v,t},
r^{\text{UP\_aFRR}}_{v,t}, r^{\text{DN\_aFRR}}_{v,t} \geq 0 ;
\label{r_UP/DN} 
\\
e^{\text{SELL\_DA}}_{v,t}/\Delta - e^{\text{BUY\_DA}}_{v,t}/\Delta 
+ r^{\text{UP\_FCR}}_{v,t} + r^{\text{UP\_aFRR}}_{v,t}
\nonumber 
\\
\leq
\text{min}( P^{\text{OBC\_MAX}}_{v}, P^{\text{CP\_MAX}}_{v,t}); 
\label{f_up}
\\
e^{\text{BUY\_DA}}_{v,t}/\Delta - e^{\text{SELL\_DA}}_{v,t}/\Delta 
+ r^{\text{DN\_FCR}}_{v,t} + r^{\text{DN\_aFRR}}_{v,t}
\nonumber 
\\
\leq \text{min}( P^{\text{OBC\_MAX}}_{v}, P^{\text{CP\_MAX}}_{v,t}); 
\label{f_dn}
\end{gather}
\begin{gather}
soe^{\text{EV}}_{v,t} = SOE^{\text{T0}} \cdot B_{v} 
+ e^{\text{OTH}}_{v,t}
+ e^{\text{ACT}}_{v,t};
\label{e_SOE} 
\\
e^{\text{OTH}}_{v,t}= \sum\nolimits_{\tau=1}^{t} \big\{
  e^{\text{BUY\_DA}}_{v,\tau} \cdot \eta^{\text{CH}}
- e^{\text{SELL\_DA}}_{v,\tau} / \eta^{\text{DCH}}
\nonumber
\\
- E^{\text{RUN}}_{v,\tau} / \eta^{\text{RUN}}  + e^{\text{FCH}}_{v,\tau} \cdot \eta^{\text{FCH}} \big\}; 
\label{sum_NRA_SOE}
\\
e^{\text{ACT}}_{v,t}= \sum\nolimits_{\tau=1}^{t} \Big\{
  \Delta \cdot \big[ \eta^{\text{CH}} \cdot (
  r^{\text{DN\_FCR}}_{v,\tau}  \cdot A^{\text{DN\_FCR}}
 \nonumber \\
+ r^{\text{DN\_aFRR}}_{v,\tau} \! \cdot \! A^{\text{DN\_aFRR}}
) \!
- \! 1/\eta^{\text{DCH}} \! \cdot \! (
r^{\text{UP\_FCR}}_{v,\tau} \! \cdot \! A^{\text{UP\_FCR}}
  \nonumber \\
+ r^{\text{UP\_aFRR}}_{v,\tau} \cdot A^{\text{UP\_aFRR}}
) \big]
\Big\}; 
\label{sum_RA_SOE}
\\
SOE^{\text{MIN}} \cdot B_{v} \leq  soe^{\text{EV}}_{v,t} 
+ e^{\text{BUY\_DA}}_{v,t+1} \cdot \eta^{\text{CH}}
\nonumber \\
- e^{\text{SELL\_DA}}_{v,t+1} \! / \eta^{\text{DCH}}
\! - \! \Lambda \! \cdot \! \Delta / \eta^{\text{DCH}} \! \cdot \! (
  r^{\text{UP\_FCR}}_{v,t+1}
\! + \! r^{\text{UP\_aFRR}}_{v,t+1})
\nonumber \\ 
- E^{\text{RUN}}_{v,t+1} / \eta^{\text{RUN}} + e^{\text{FCH}}_{v,t+1} \cdot \eta^{\text{FCH}} 
\quad \forall t \in \mathcal{T}_{(t\neq N_t)};
\label{soe_MINR}
\\
SOE^{\text{MAX}} \cdot B_{v} \geq 
soe^{\text{EV}}_{v,t}
+ e^{\text{BUY\_DA}}_{v,t+1} \cdot \eta^{\text{CH}}
\nonumber \\
- e^{\text{SELL\_DA}}_{v,t+1} \! / \eta^{\text{DCH}}
\!+\! \Lambda \!\cdot\! \Delta \!\cdot\! \eta^{\text{CH}} \!\cdot\! (
  r^{\text{DN\_FCR}}_{v,t+1}
\!+\! r^{\text{DN\_aFRR}}_{v,t+1})
\nonumber \\
- E^{\text{RUN}}_{v,t+1} / \eta^{\text{RUN}}  + e^{\text{FCH}}_{v,t+1} \cdot \eta^{\text{FCH}} 
\quad \forall t \in \mathcal{T}_{(t\neq N_t)};
 \label{soe_MAXR}
\\
SOE^{\text{T0}} \! \cdot B_{v} \leq soe^{\text{EV}}_{v,t} \leq SOE^{\text{MAX}} \! \cdot B_{v} \quad \text{for} \  t=N_{t}; 
\label{soe_T24}
\\
0 \leq e^{\text{FCH}}_{v,t} \leq P^{\text{FCH\_MAX}} \cdot \Delta;
\label{e_FCH_CP}
\\
e^{\text{DEG}}_{v,t} =
  e^{\text{SELL\_DA}}_{v,t} / \eta^{\text{DCH}}
-
e^{\text{BUY\_DA}}_{v,t} \cdot \eta^{\text{CH}}
\nonumber \\
 + \Delta / \eta^{\text{DCH}} \!\cdot\!
 ( r^{\text{UP\_FCR}}_{v,t} \!\cdot\! A^{\text{UP\_FCR}}
\!+\! r^{\text{UP\_aFRR}}_{v,t} \!\cdot\! A^{\text{UP\_aFRR}})
\nonumber \\
-
\Delta \!\cdot\! \eta^{\text{CH}} \!\cdot\!
 (r^{\text{DN\_FCR}}_{v,t} \!\cdot\! A^{\text{DN\_FCR}}
\!+ \!r^{\text{DN\_aFRR}}_{v,t} \!\cdot\! A^{\text{DN\_aFRR}}
);
%\Big]
\label{DEG}  \\
c^{\text{DEG}}_{v,t} >= 0; \label{DEG_3}
\\
%c^{\text{DEG}}_{v,t} >= C^{\text{B}}_{v} \cdot (-D^{\text{B}}_{1}
c^{\text{DEG}}_{v,t} \cdot B_{v} >= C^{\text{B}}_{v} \cdot \big[-D^{\text{B}}_{1} \cdot B_{v} 
\nonumber 
\\
%+ D^{\text{B}}_{2} \cdot \frac {e^{\text{DEG}}_{v,t}} {B_{v}} 
+ D^{\text{B}}_{2} \cdot e^{\text{DEG}}_{v,t}
%+ D^{\text{B}}_{3} \cdot \frac{{B_{v}}-soe^{\text{EV}}_{v,t}} {B_{v}} ); 
+ D^{\text{B}}_{3} \cdot (B_{v}-soe^{\text{EV}}_{v,t})\big]; 
\label{DEG_1}  \\
%c^{\text{DEG}}_{v,t} >=  C^{\text{B}}_{v} \cdot D^{\text{B}}_{4} \cdot \frac {e^{\text{DEG}}_{v,t}} {B_{v}}. \label{DEG_2}
c^{\text{DEG}}_{v,t}\cdot B_{v}>=  C^{\text{B}}_{v} \cdot D^{\text{B}}_{4} \cdot e^{\text{DEG}}_{v,t}. \label{DEG_2}
\end{gather}

Eq. \eqref{OF_NRA} sums the DA market costs (consisting of energy purchase and sell as well as up and down FCR and aFRR capacity reservation), battery degradation and fast charging cost. On the other hand, eq. \eqref{OF_RA} contains only the RA costs, where up RA is remunerated by the system operator (hence the minus sign), while the down RA must be payed to the system operator (hence the plus sign).
Eq. \eqref{r_UP/DN} sets the six market bidding variables as nonnegative. Eqs. \eqref{f_up} and \eqref{f_dn} limit the total charging/discharging power available for bidding to the minimum of the On-Board Charger (OBC) capacity and the Charging Point (CP) capacity. %The CP capacity has been prepared as energy that an EV can maximally charge/discharge in one time-step and it originates from the input format of the EV driving/parking behaviour explained in Subsection \ref{sec:inputs}. 
In other words, \eqref{f_up} and \eqref{f_dn} allocate the available power to the six market bidding variables appearing in these constraints.

Eq. \eqref{e_SOE} calculates the current SOE based on the initial SOE and the energy charged/discharged to/from the EV's battery until the time-step $t$. The energy can be charged/discharged by the means of DA market trading, driving, fast charging \eqref{sum_NRA_SOE} and RA \eqref{sum_RA_SOE}. Energy consumed for driving in one time-step originates from the input data of the EV driving/parking behaviour explained in Subsection \ref{sec:inputs}. The amount of energy injected/extracted due to RA is modeled using parameters $A^{\text{\{UP/DN\}\_\{FCR/aFRR\}}} \! \in \! [0,1]$.
%E.g. term $-r^{\text{UP\_FCR}}_{v,t} \cdot A^{\text{UP\_FCR}} \cdot \Delta$ refers to activated up FCR reserve which can be an increase in discharging or decrease in charging. 
To make sure that the EV batteries will be able to deliver the required reserves, their capacity is limited in eqs. \eqref{soe_MINR} and \eqref{soe_MAXR} assuming a their full activation. Eq. \eqref{soe_MINR} acts as the lower bound to the SOE, where only the full up RA is considered. Similarly, eq. \eqref{soe_MAXR} acts as the upper bound to the SOE considering the full down RA. These two equations ensure that the SOE is feasible in each time step even for the full RA. Eqs. \eqref{soe_MINR} and \eqref{soe_MAXR} are applied up to timestep $t<N_t$. In the last timestep the conventional SOE preservation eq. \eqref{soe_T24} is applied.
%They replace the conventional constraints on SOE ($SOE^{\text{MIN}} \cdot B_{v} \leq soe^{\text{EV}}_{v,t} \leq SOE^{\text{MAX}} \cdot B_{v}$). To bound the market variables in time-step $(t+1)$ at $t=N_t$, in both equations, the conventional SOE constraints from the previous sentence are used. These equations take care of the current SOE bounds, but still due to uncertain activation in prior time-steps $soe^{\text{EV}}_{v,t}$ could carry errors.
The fast-charging limit is enforced in \eqref{e_FCH_CP}. %Fast charging is utilized only if there is insufficient energy to complete the trip, i.e. variable $e^\mathrm{FCH}_{v,t}$ ensures the model feasibility. 

Li-ion batteries are prone to degradation, especially when cycled often. Incorporating degradation cost in the OF may reduce the battery charging/discharging actions not related to driving. The degradation is taken into account when discharging in the DA market or through RA in eq. \eqref{DEG}. Eq. \eqref{DEG_3} sets  $c^{\text{DEG}}_{v,t}$ as a positive variable.  Eqs. \eqref{DEG_1} and \eqref{DEG_2} bound and calculate the V2G discharging degradation cost. Those constraints are a linearized form of the degradation model from \cite{Ortega}.
The above constraints \eqref{r_UP/DN}--\eqref{DEG_2} apply for $\forall v \in \mathcal{V}$ and $\forall t \in \mathcal{T}$, except for eqs. \eqref{soe_MINR}--\eqref{soe_MAXR}, which are not valid in the last period, and for \eqref{soe_T24}, which is valid in the last period only.
\vspace{-4mm}
\subsection{Stochastic Model -- SM} \label{sec:stoc}
\vspace{-2mm}
\renewcommand{\theequation}{S-\arabic{equation}}
\setcounter{equation}{0}

The stochastic model (SM) differs from the DM in the definition of the RA ratio. While the DM uses one value for all timesteps, the SM uses the RA ratio as a time-dependable parameter taken from the historic data using several scenarios (details on how this data is used is explained in Section \ref{sec:RAM}). The RA parameters $A^{\text{\{UP/DN\}\_\{FCR/aFRR\}}}_{s,t}$ and the associated variables $g^{\text{RA}}_{s}, h^{\text{RA}}_{s}, soe^{\text{EV}}_{s,v,t},  e^{\text{DEG}}_{s,v,t}, c^{\text{DEG}}_{s,v,t}$ gain an additional index $s$. The model is accordingly reformulated using eqs. \eqref{OF_STOC}--\eqref{STOC_s}. The OF of the SM is \eqref{OF_STOC}. Eqs. in \eqref{Stoc_nos} are identical as in the DM, while the SM instances of eqs. \eqref{STOC_As} and \eqref{STOC_s} include additional index $\forall s \in \mathcal{S}$. %Eqs. \eqref{STOC_As} are those directly containing activation parameters, while eqs. \eqref{STOC_s} contain variables with new index $s$.
Objective function:
%With scenarios
\vspace{-2mm}
\begin{gather} 
\min_{\Xi_{\mathcal{O}}} \ [P_{s} \cdot \sum\nolimits_{s=1}^{N_{s}} ( c^{\text{OTH}}_{s}
+c^{\text{ACT}}_{s}
)];
\label{OF_STOC}
\end{gather}
\vspace{-3mm}
subject to:
\vspace{-2mm}
\begin{gather}
\eqref{r_UP/DN} - \eqref{f_dn};
\label{Stoc_nos}\\
\eqref{OF_RA}, \eqref{sum_RA_SOE},  \eqref{DEG}, \quad \forall s \in \mathcal{S};
\label{STOC_As} \\
\eqref{OF_NRA}, \eqref{e_SOE} - \eqref{sum_NRA_SOE}, \eqref{soe_MINR} - \eqref{e_FCH_CP},
\nonumber \\
\eqref{DEG_3} - \eqref{DEG_2}, \quad \forall s \in \mathcal{S}.
\label{STOC_s}  
\end{gather}
\vspace{-7mm}
\subsection{Robust Model -- RM} \label{sec:rob}
\vspace{-1mm}
In the robust model (RM) the RA quantities are uncertain parameters ($a^{\text{UP\_FCR}}_{v, \tau, t}, a^{\text{DN\_FCR}}_{v, \tau, t}, a^{\text{UP\_aFRR}}_{v, \tau, t}, a^{\text{DN\_aFRR}}_{v, \tau, t}$) whose boundaries are defined by the uncertainty set (US) defined in eqs. \eqref{pos_US}--\eqref{cons_us} with parameters resulting from the probabilistic analysis in Section \ref{sec:RAM}. OF \eqref{OF_z} of the RM minimizes the total operating costs for the worst-case RA, i.e. maximizing the RA throughout the day, \eqref{z_cons}--\eqref{SOE_nt_rob}. Such min-max structure cannot be solved directly and an appropriate RM reformulation presented in \cite{Morales} was used.

The initial formulation of the RM is presented in Section \ref{sec:RIF}. Using the duality theory, subproblems \eqref{z_cons}--\eqref{SOE_nt_rob} are recast as duals and merged with the rest of the constraints creating the final problem presented in Section \ref{sec:final}.
\subsubsection{Initial Formulation} \label{sec:RIF}
\hspace*{\fill}
\renewcommand{\theequation}{RI-\arabic{equation}}
\setcounter{equation}{0}

Objective function:
\vspace{-2mm}
\begin{gather} 
\min_{\Xi_{\mathcal{O}}} \quad (z)
\label{OF_z}
\end{gather}
\vspace{-2mm}
is subject to:
\vspace{-1mm}
\begin{gather}
\eqref{OF_NRA}, \eqref{r_UP/DN} - \eqref{f_dn}, \eqref{sum_NRA_SOE}, \eqref{e_FCH_CP} - \eqref{DEG_2};
\label{Rob_Ns} \\
\eqref{OF_RA}, \eqref{sum_RA_SOE};
\label{Rob_As} \\
\max_{\Xi_{\mathcal{A}}} \ (c^{\text{ACT}})
\leq z - c^{\text{OTH}}; 
\label{z_cons} \\
\max_{\Xi_{\mathcal{A}}} 
(-e^{\text{ACT}}_{v,t})
\leq 
B_{v} \cdot (SOE^{\text{T0}} - SOE^{\text{MIN}}) 
 \nonumber  \\
- \Lambda \cdot \Delta / \eta^{\text{DCH}} \cdot (
  r^{\text{UP\_FCR}}_{v,t+1}
+ r^{\text{UP\_aFRR}}_{v,t+1})
+ e^{\text{OTH}}_{v,t+1};
\label{SOE_min_rob}
\\
\max_{\Xi_{\mathcal{A}}} (e^{\text{ACT}}_{v,t})
\leq 
B_{v} \cdot (SOE^{\text{MAX}} - SOE^{\text{T0}}) 
\nonumber
\\
- \Lambda \cdot \Delta \cdot \eta^{\text{CH}} \cdot (
  r^{\text{DN\_FCR}}_{v,t+1}
+ r^{\text{DN\_aFRR}}_{v,t+1})
- e^{\text{OTH}}_{v,t+1};
\label{SOE_max_rob} \\
\max_{\Xi_{\mathcal{A}}} \
(e^{\text{ACT}}_{v,t})
\leq 
B_{v} \cdot (SOE^{\text{MAX}} - SOE^{\text{T0}}) 
- e^{\text{OTH}}_{v,t}; 
\label{SOE_nt_rob_geq}
\\
\max_{\Xi_{\mathcal{A}}} \
(-e^{\text{ACT}}_{v,t})
\leq 
 e^{\text{OTH}}_{v,t}, \quad \text{for} \quad t=N_{t}; 
\label{SOE_nt_rob}
\end{gather}
OF of the RM \eqref{OF_z} minimizes the total cost of an EV fleet. Eqs. \eqref{Rob_Ns} are the same as in the DM, while the uncertain parameters are located in the constraints containing the RA variables, grouped under eq. \eqref{Rob_As}. Eqs. \eqref{Rob_As} are similar as in the DM, but the fixed RA parameter is replaced with an uncertain one. OF and each constraint containing the terms with uncertain parameters from \eqref{Rob_As} are observed as independent maximization subproblems \eqref{z_cons}--\eqref{SOE_nt_rob}. OF of the DM eqref{OF} is reformulated to its robust counterpart presented in eqs. \eqref{OF_z} and \eqref{z_cons}. Eqs. \eqref{soe_MINR}--\eqref{soe_T24} bounding the SOE are adequatly reformulated to eqs. \eqref{SOE_min_rob}--\eqref{SOE_nt_rob}. %Note that the SOE balance equation \eqref{e_SOE} is already incorporated in aforementioned eqs.
\renewcommand{\theequation}{US-\arabic{equation}}
\setcounter{equation}{0}

Subproblems defined by eqs. \eqref{z_cons}--\eqref{SOE_nt_rob} are valid $\forall \ (a^{\text{UP\_FCR}}_{v, \tau, t}, a^{\text{DN\_FCR}}_{v, \tau, t}, a^{\text{UP\_aFRR}}_{v, \tau, t}, a^{\text{DN\_aFRR}}_{v, \tau, t}) \in \mathcal{A}$, where $\mathcal{A}$ is the following US:
\vspace{-2mm}
\begin{gather}
\mathcal{A} = \big\{
a^{\text{UP\_FCR}}_{v, \tau, t}, a^{\text{DN\_FCR}}_{v, \tau, t}, a^{\text{UP\_aFRR}}_{v,\tau, t}, a^{\text{DN\_aFRR}}_{v, \tau, t}
\quad | 
\nonumber \\
a^{\text{UP\_FCR}}_{v, \tau, t}, a^{\text{DN\_FCR}}_{v, \tau, t}, a^{\text{UP\_aFRR}}_{v, \tau, t}, a^{\text{DN\_aFRR}}_{v, \tau, t} \geq 0;
\label{pos_US}
\\
a^{\text{UP\_FCR}}_{v, \tau, t} + a^{\text{DN\_FCR}}_{v, \tau, t} \leq A^{\text{MAX\_FCR}} 
:\omega^{\eqref{SOE_min_rob}\_\text{FCR}}_{v, \tau, t};
\label{AMAX_FCR_CN}
\\
\sum\nolimits_{\tau=1}^{t} \! a^{\text{UP\_FCR}}_{v, \tau, t} \!\leq\!
(\Gamma^{\text{UP\_FCR}} \!- \!1) \!\cdot \!I_{t} \!+ \!1
:\mu^{\eqref{SOE_min_rob}\_\text{FCR}}_{v, t};
\label{UP_FCR_MAX_CN}
\\
\sum\nolimits_{\tau=1}^{t} a^{\text{UP\_FCR}}_{v,\tau, t} \geq \Upsilon^{\text{UP\_FCR}} \cdot I_{t}
:\nu^{\eqref{SOE_min_rob}\_\text{FCR}}_{v,t};
\label{UP_FCR_MIN_CN} 
\\
\sum\nolimits_{\tau=1}^{t} \!a^{\text{DN\_FCR}}_{v,\tau, t} \!\leq \!
(\Gamma^{\text{DN\_FCR}} \!-\! 1) \!\cdot\! I_{t} \!+ \!1
:\psi^{\eqref{SOE_min_rob}\_\text{FCR}}_{v,t};
\label{DN_FCR_MAX_CN} 
\\
\sum\nolimits_{\tau=1}^{t} a^{\text{DN\_FCR}}_{v,\tau, t} \geq \Upsilon^{\text{DN\_FCR}} \cdot I_{t}
:\chi^{\eqref{SOE_min_rob}\_\text{FCR}}_{v,t};
\label{DN_FCR_MIN_CN} 
\\
\text{\eqref{AMAX_FCR_CN}--\eqref{DN_FCR_MIN_CN} are analogous for aFRR;}
\label{aFRR_us_CN} \\
\text{\eqref{AMAX_FCR_CN}--\eqref{aFRR_us_CN} are similar for \eqref{z_cons}--\eqref{SOE_nt_rob}.}
\big\}
\label{cons_us}
\end{gather}
The US presented in eqs. \eqref{pos_US}--\eqref{DN_FCR_MIN_CN} applies to the \textit{FCR} service and for the subproblem stated in \eqref{SOE_min_rob}. The uncertainty set \eqref{aFRR_us_CN} spreads it over the \textit{aFRR} service and \eqref{cons_us} to other robust subproblems. For constraints \eqref{SOE_min_rob} and \eqref{SOE_max_rob} the US is applied up to a specific time-step $t$ and for a specific EV $v$. The US equations, uncertain parameters and dual variables for \eqref{SOE_max_rob} are the same as for \eqref{SOE_min_rob}.

The US for OF \eqref{z_cons} is applied over the entire horizon $[1,N_{t}]$ and entire EV fleet $[1, N_{v}]$, whereas in \eqref{SOE_nt_rob_geq} and \eqref{SOE_nt_rob} it is applied up to the last time-step $N_t$ (mathematically it is the same as for the \eqref{z_cons}) and for a specific EV $v$. 
Following this reasoning, index $\tau$ is omitted in the dual variables, uncertain parameters and the summation terms in \eqref{UP_FCR_MAX_CN}--\eqref{DN_FCR_MIN_CN} (the sum goes from $t \in [1,N_{t}]$) for \eqref{z_cons}, \eqref{SOE_nt_rob_geq} and \eqref{SOE_nt_rob}. For \eqref{z_cons}, index $v$ is also omitted.
Therefore, for OF \eqref{z_cons} the dual variable of constraint \eqref{AMAX_FCR_CN} is indexed only with $t$ (i.e. $\omega^{\eqref{z_cons}}_{t}$) and the dual variables of the rest of the constraints \eqref{UP_FCR_MAX_CN}--\eqref{DN_FCR_MIN_CN} do not have an index (e.g. $\mu^{\eqref{z_cons}}$). Dual variables for \eqref{SOE_nt_rob_geq} and \eqref{SOE_nt_rob} are the same as for \eqref{z_cons} but they include index $v$ (e.g. $\omega^{\eqref{SOE_nt_rob}}_{t,v}$, $\mu^{\eqref{SOE_nt_rob}}_{v}$).

For OF \eqref{z_cons}, \eqref{SOE_nt_rob_geq}--\eqref{SOE_nt_rob}, parameter $I_{t}$ is always equal to $1$, while for constraints \eqref{SOE_min_rob}--\eqref{SOE_max_rob} it represents a uniformly distributed range $\in [0,1]$ for index \textit{t}  $\in [1,N_{t}]$. Parameter $I_{t}$ is designed to partition the min ($\Upsilon$)/max ($\Gamma$) total activated reserve in time-frames shorter than $N_{t}$.

\subsubsection{Final Formulation} \label{sec:final}
\hspace*{\fill}

\renewcommand{\theequation}{RF-\arabic{equation}}
\setcounter{equation}{0}

Objective function:
\vspace{-3mm}
\begin{gather} 
\min_{\Xi_{\mathcal{O}}} \quad (z)
\label{OF_z2}
\end{gather}
\vspace{-3mm}
subject to:
\vspace{-1mm}
%%%%%%%%%%%Ovdje počinje za RI5
%
\begin{gather}
\eqref{OF_NRA}, \eqref{r_UP/DN} - \eqref{f_dn}, \eqref{sum_NRA_SOE}, \eqref{e_FCH_CP} - \eqref{DEG_2}; \\
\eqref{OF_RA}, \eqref{sum_RA_SOE};
\label{Rob_As2}
\\
G^{\text{\{UP/DN\}\_\{FCR/aFRR\}}}_{t} \!=
(\Gamma^{\text{\{UP/DN\}\_\{FCR/aFRR\}}}\text{-}1) \cdot I_{t} \text{+}1
\label{f}
\\
A^{\text{MAX\_FCR}} \cdot \sum\nolimits_{\tau=1}^{t} \omega^{\eqref{SOE_min_rob}\_\text{FCR}}_{v, \tau, t} 
\nonumber \\
+ G^{\text{UP\_FCR}}_{t} \!\cdot \!\mu^{\eqref{SOE_min_rob}\_\text{FCR}}_{v, t}
\!+ \!\Upsilon^{\text{UP\_FCR}} \!\cdot\! I_{t} \!\cdot \!
\nu^{\eqref{SOE_min_rob}\_\text{FCR}}_{v, t}
\nonumber
\\
+ G^{\text{DN\_FCR}}_{t} \!\cdot\! \psi^{\eqref{SOE_min_rob}\_\text{FCR}}_{v, t}
\!+\! \Upsilon^{\text{DN\_FCR}} \!\cdot \!I_{t}\! \cdot\!
\chi^{\eqref{SOE_min_rob}\_\text{FCR}}_{v, t}
\nonumber  
\\
A^{\text{MAX\_aFRR}} \cdot \sum\nolimits_{\tau=1}^{t} \omega^{\eqref{SOE_min_rob}\_\text{aFRR}}_{v, \tau, t} 
\nonumber \\
+ G^{\text{UP\_aFRR}}_{t} \!\cdot \!\mu^{\eqref{SOE_min_rob}\_\text{aFRR}}_{v, t}
\!+ \!\Upsilon^{\text{UP\_aFRR}} \!\cdot \!I_{t} \!\cdot \!
\nu^{\eqref{SOE_min_rob}\_\text{aFRR}}_{v, t}
\nonumber \\
+ G^{\text{DN\_aFRR}}_{t} \!\cdot\! \psi^{\eqref{SOE_min_rob}\_\text{aFRR}}_{v, t}
\!+ \!\Upsilon^{\text{DN\_aFRR}} \!\cdot\! I_{t}\! \cdot\!
\chi^{\eqref{SOE_min_rob}\_\text{aFRR}}_{v, t}
\nonumber
\\
\leq 
B_{v} \cdot (SOE^{\text{T0}} - SOE^{\text{MIN}}) 
\nonumber
\\
- \Lambda \cdot \Delta / \eta^{\text{DCH}} \cdot (
  r^{\text{UP\_FCR}}_{v,t+1}
+ r^{\text{UP\_aFRR}}_{v,t+1})
+ h^{\text{NRA}}_{v,t+1};
\label{e_CH/DCH_uncertain2}
\\
\omega^{\eqref{SOE_min_rob}\_\text{FCR}}_{v, \tau, t} + \mu^{\eqref{SOE_min_rob}\_\text{FCR}}_{v, t} + 
\nu^{\eqref{SOE_min_rob}\_\text{FCR}}_{v, t} \geq
\nonumber \\
\Delta / \eta^{DCH} \cdot  r^{\text{UP\_FCR}}_{v,\tau} \quad
:a^{\text{UP\_FCR}}_{v, \tau, t};
\label{CN_aUP_FCR}
\\
\omega^{\eqref{SOE_min_rob}\_\text{FCR}}_{v, \tau, t} + \psi^{\eqref{SOE_min_rob}\_\text{FCR}}_{v, t} + 
\chi^{\eqref{SOE_min_rob}\_\text{FCR}}_{v, t} \geq \nonumber \\
-\Delta \cdot \eta^{CH} \cdot r^{\text{DN\_FCR}}_{v,\tau} \quad
:a^{\text{DN\_FCR}}_{v, \tau, t};
\label{CN_aDN_FCR}
\\
\omega^{\eqref{SOE_min_rob}\_\text{FCR}}_{v, \tau, t}, \mu^{\eqref{SOE_min_rob}\_\text{FCR}}_{v, t}, \psi^{\eqref{SOE_min_rob}\_\text{FCR}}_{v, t} \geq 0;
\label{CN_bound_FCR1}
\\
\quad \nu^{\eqref{SOE_min_rob}\_\text{FCR}}_{v, t}, \chi^{\eqref{SOE_min_rob}\_\text{FCR}}_{v, t} \leq 0;
\label{CN_bound_FCR2} 
\\
\text{\eqref{CN_aUP_FCR}--\eqref{CN_bound_FCR2} are analogous for FRR;}
\label{aFRR_cons} \\
\text{\eqref{f}--\eqref{aFRR_cons} are analogous for \eqref{z_cons}--\eqref{SOE_nt_rob}}.
\label{RI_full}
\end{gather}
Eqs. \eqref{OF_z2}--\eqref{Rob_As2} are the same as in the RI formulation. Auxiliary eq. \eqref{f} is  used to shorten the notation.  Eqs. \eqref{e_CH/DCH_uncertain2}--\eqref{CN_bound_FCR2} represent strong duality and dual constraints (for FCR) of the subproblem stated in \eqref{SOE_min_rob}. Eq. \eqref{e_CH/DCH_uncertain2} applies $\forall t \in \mathcal{T}$ and $\forall v \in \mathcal{V}$, while eqs. \eqref{CN_aUP_FCR}--\eqref{CN_bound_FCR2} additionally apply $\forall \tau \in \mathcal{T}, \tau < t$.  Eq. \eqref{aFRR_cons} spreads the problem over aFRR dual constraints and eq. \eqref{RI_full} spreads it on subproblems \eqref{z_cons}, \eqref{SOE_max_rob}--\eqref{SOE_nt_rob}. The left-hand side in eqs. \eqref{e_CH/DCH_uncertain2}--\eqref{CN_aDN_FCR} is the same for subproblem \eqref{SOE_max_rob} (only the superscript of dual variables changes), whereas for subproblems \eqref{z_cons}, \eqref{SOE_nt_rob_geq} and \eqref{SOE_nt_rob} the indices must be aligned to those of the US dual variables (the sum in \eqref{e_CH/DCH_uncertain2} goes from $t \in [1,N_{t}]$). Constraints for \eqref{z_cons} are simplified versions of \eqref{e_CH/DCH_uncertain2}--\eqref{aFRR_cons} as they apply over all $\mathcal{T}$ and $\mathcal{V}$. Eqs. for subproblems \eqref{SOE_nt_rob_geq} and \eqref{SOE_nt_rob} retain index $\forall v \in \mathcal{V}$.
The right-hand sides in eqs. \eqref{e_CH/DCH_uncertain2}--\eqref{CN_aDN_FCR} must be aligned with cost coefficients on the left-hand side and limits on the right-hand side of \eqref{z_cons}--\eqref{SOE_nt_rob}.
\vspace{-2mm}
\section{Case Studies} \label{sec:case}
The developed models were deployed to obtain the DA schedules. The quality of the obtained schedules was assessed by running them against one hundred historical RA scenarios. 
%There are three major sets of inputs: market prices, EV data and reserve activations. Only reserve activations are modeled as random parameters while the rest are kept fixed. 
Three models are tested: the DM with average annual RA, the SM with 10 ex-ante RA scenarios and the RM with the US based on the real balancing data. 
\vspace{-2mm}
\subsection{Reserve Activation -- RA} \label{sec:RAM}
\begin{figure} 
\begin{subfigure}{.5\textwidth}
    \centering
    \includegraphics[trim=15 30 55 70, clip, width=0.7\linewidth]{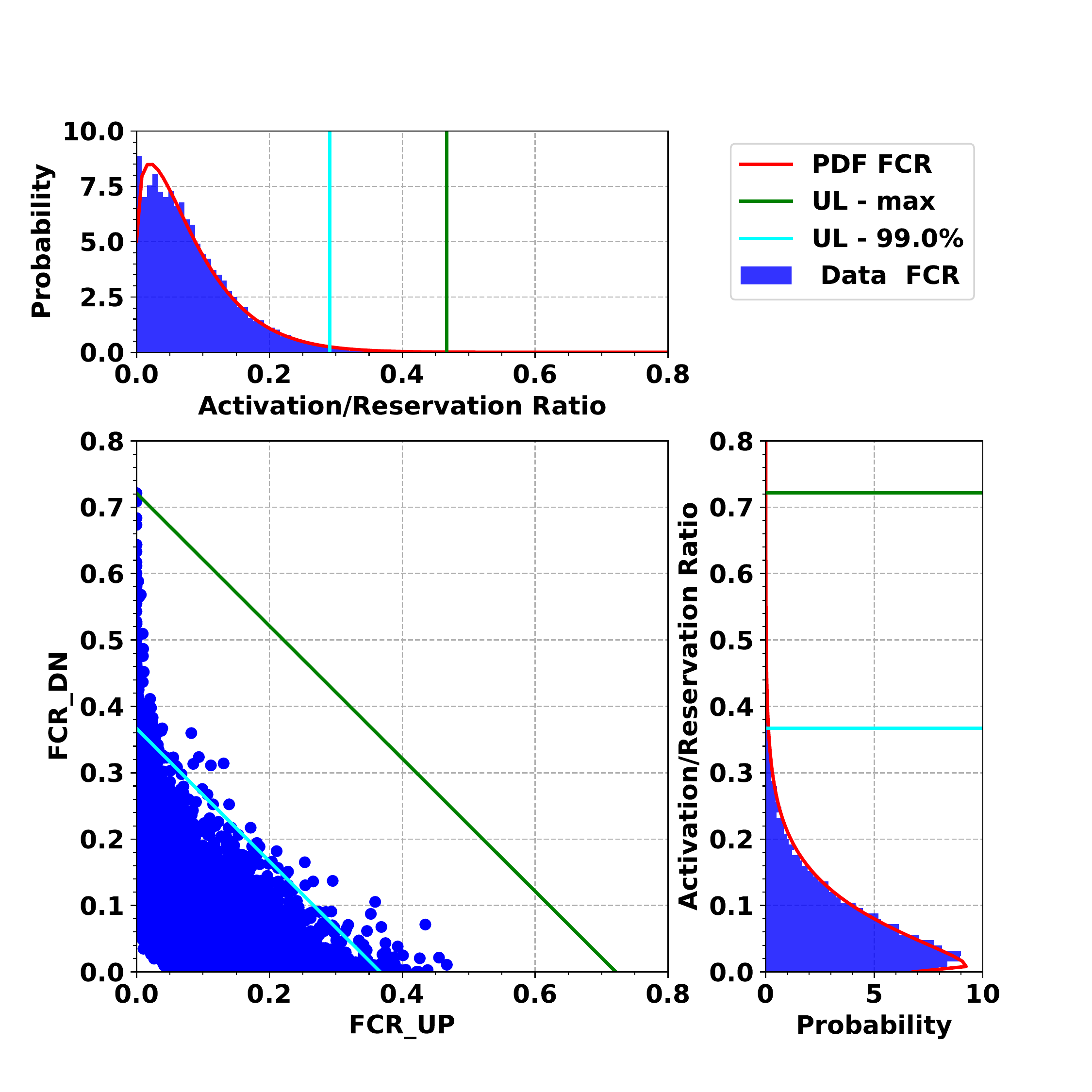}
    \captionsetup{font=footnotesize, labelsep=period, justification   = raggedright}
    \caption{FCR Activation Ratios}
    \label{fig:FCR_prob}
\end{subfigure}
\begin{subfigure}{.5\textwidth}
    \centering
    \includegraphics[trim=15 30 55 70, clip, width=0.7\linewidth]{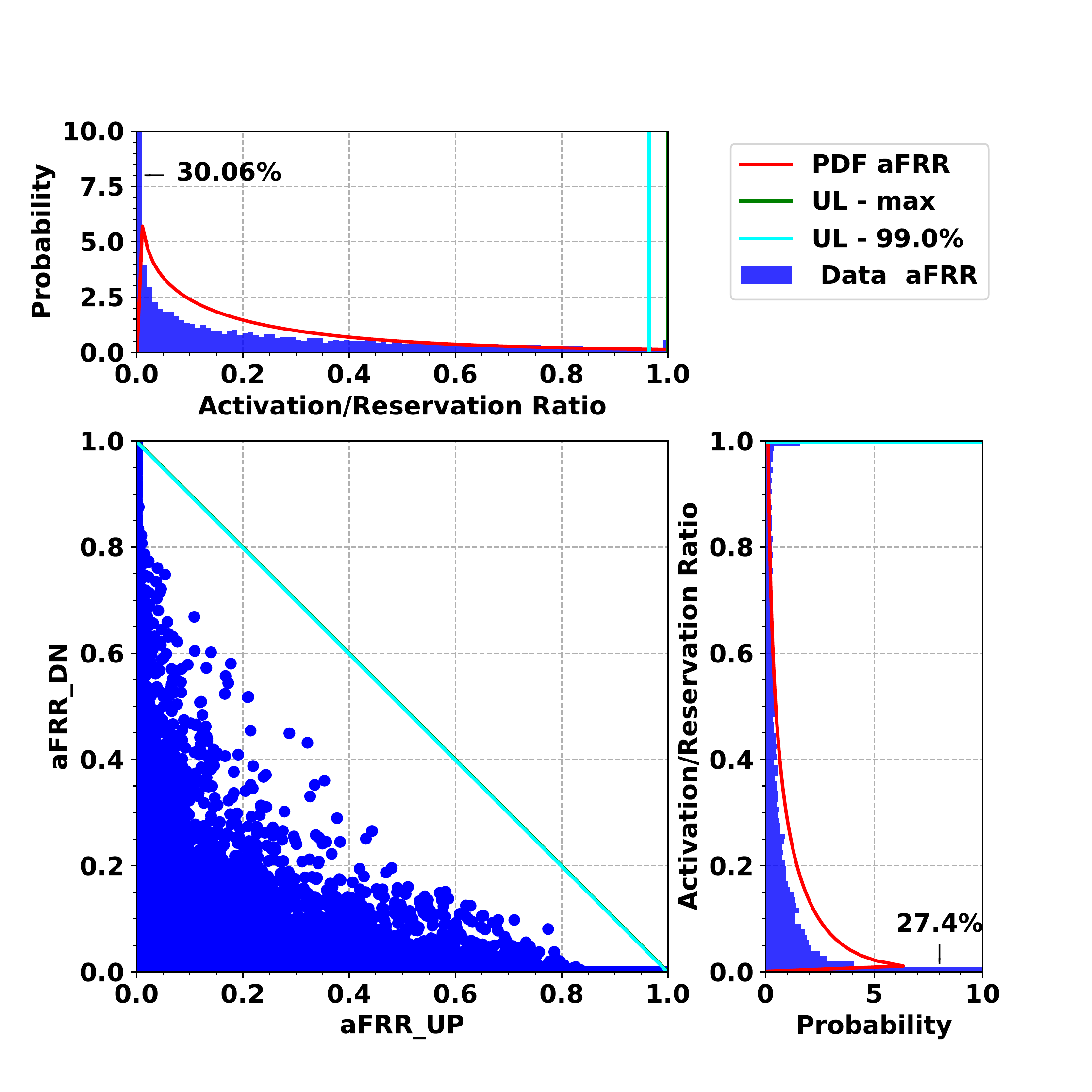}
    \captionsetup{font=footnotesize, labelsep=period, justification   = raggedright}
    \caption{aFRR Activation Ratios}
    \label{fig:FRR_prob}
\end{subfigure}
\captionsetup{font=footnotesize, labelsep=period, justification   = raggedright}
    \caption{Activation Ratios -- Historical Data}
    \label{fig:prob}
    \vspace{-4mm}
\end{figure}

\renewcommand{\theequation}{A-\arabic{equation}}
\setcounter{equation}{0}
The core input data to calculate the RA ratios are \textit{FCR/aFRR Accepted Reserve Capacity (ARC)} bids and \textit{FCR/aFRR Activated Reserve Energies (ARE)} taken from the RTE for the year 2018 \cite{RTE}. For each half-hourly period (and each reserve type and direction) the RA quantities are calculated as:
\begin{gather}
\bigg(RA\_ratio_{t}  = 
\frac {ARE_{t}} {ARC_{t}
\cdot \Delta}\bigg)^{\text{\{UP/DN\}\_\{FCR/aFRR\}}} 
\label{A_DN_RES}
\end{gather}
Using the calculated half-hourly ratios, a statistical analyses was performed to gain the required input data. Visualization of the obtained results for FCR and aFRR are presented in Figure \ref{fig:prob}. These data are used to determine the average values for the DM, to select scenarios for the SM, and to obtain inputs for the US of the RM. Also, it will be used to select RA scenarios for the ex-post validation of the models.
%left bottom right top
%Need to increase the letters

Probability distribution functions (PDF) for up and down reserves are very similar for both the FCR and the aFRR. The down reserve has a slightly higher annual average value for both the FCR and the aFRR. Also, average values for both UP and DN are much higher for aFRR than for FCR. The activation values used in the DM are shown in Table \ref{tab:RA} in the second column.
%: $A^{\text{UP\_FCR}} \!=\! 0.082$, $A^{\text{DN\_FCR}} \! = \! 0.085$, $A^{\text{UP\_aFRR}} \!=\! 0.198$, and $A^{\text{DN\_aFRR}} \!=\! 0.218$.
Scenarios for the SM are taken as the realized RA ratios for each reserve type and direction from Jan. 11-20, 2018 (10 in total). Ex-post validation is performed on scenarios from April 1, 2018 to July 9, 2018 (100 in total).

For the US of the RM two values are required, the maximum achieved RA ratios in each time-step and the overall daily activated energy. %. The maximum RA ratios represent the power that could be activated in one time-step, while the summations represent the total energy that could be yielded through RA in one day. %In this paper, two sets of US parameters were used: one for subproblem \eqref{z_cons} and one for subproblems \eqref{SOE_min_rob} -- \eqref{SOE_nt_rob}. This division is applied because the case studies are envisioned to prove that SOE limits can be preserved efficiently with RM and the robust modeling of OF is an addition which can improve the results. Therefore, the parameters for subproblem \eqref{z_cons} are mild, whereas for the subproblems \eqref{SOE_min_rob} -- \eqref{SOE_nt_rob} they are harsh.
The RA ratios for UP and DN reserves are dependent variables where a high RA value of one direction entails a low RA value of the other direction and vice versa (valid for both the FCR and the aFRR). This statement is modeled in \eqref{AMAX_FCR_CN} and shown as green (max) and cyan (99\%) lines on graphs in Figures \ref{fig:FCR_prob} and \ref{fig:FRR_prob}. This also bounds the individual RA ratios of the up and down reserves. 
%FCR UP and DN are never zero at the same time (there is always slight imbalance in one direction). On the contrary, aFRR UP and DN could be zero at the same time. Due to these reasons the upper and lower boundary was set as green and orange curve, respectively, on scatter plots of figures \ref{fig:FCR_prob} and \ref{fig:FRR_prob}. The equation for those boundary curves are:
The FCR RA ratio never reaches $1$; the maximum value for UP RA ratio is $0.47$ and for DN RA ratio $0.73$. In 99\% of the points the UP and DN RA ratios are lower or equal to $0.29$ and $0.36$, respectively.
High aFRR RA ratios are also rare, but more frequent than in the case of FCR so the 99\% lines in Figure \ref{fig:FRR_prob} are at $0.96$ and $1$ for UP and DN RA ratios, respectively. The parameters used in the case study are shown in Table \ref{tab:RA}. Those in the third column are used for subproblem \eqref{z_cons} and those in the fifth column are used for subproblems \eqref{SOE_min_rob}--\eqref{SOE_nt_rob}.
%$A^{\text{MAX\_FCR}}\! = \!0.36$ for subproblems \eqref{SOE_min_rob}--\eqref{SOE_nt_rob} and $A^{\text{MAX\_FCR}}\! =\! 0.73$ for \eqref{z_cons}. For all subproblems $A^{\text{MAX\_aFRR}} \!=\! 1$ was used.
%
The daily sums of the RA ratios throughout the year are always within certain upper and lower limits, which is modeled in the US constraints \eqref{UP_FCR_MAX_CN}--\eqref{DN_FCR_MIN_CN}.
The median of the daily sums for each of the reserve and direction are shown in Table \ref{tab:RA} in column four. This data is used as both $\Upsilon$ and $\Gamma$ parameters for subproblem \eqref{z_cons}.
The minimum and maximum of the daily sums for each of the reserve and direction are shown in Table \ref{tab:RA} in columns six and seven, respectively. This data is used as $\Upsilon$ and $\Gamma$ parameters for subproblems \eqref{SOE_min_rob}--\eqref{SOE_nt_rob}. 
%
%The minimum and maximum of the daily sums of the FCR UP RA ratios are $0.93$ and $8.21$, of FCR DN RA ratios are $1.10$ and $16.64$, of aFRR UP RA ratios are $2.10$ and $21.31$, and of aFRR DN RA ratios are $2.62$ and $20.51$. The median of daily sums of FCR UP RA raios is $4$, of FCR DN RA ratios are $3.8$, of aFRR UP RA ratios are $9.13$, and of aFRR DN RA ratios are $9.70$. In this paper, we used the min ($\Upsilon$s) and max ($\Gamma$s) values for the US parameters of subproblems \eqref{SOE_min_rob}--\eqref{SOE_nt_rob} and the median for the subproblem \eqref{z_cons} ($\Gamma$s are equal to $\Upsilon$s in this case).

\setlength{\tabcolsep}{4.5pt} %
\begin{table}[!b]
\vspace{-3mm}
\centering
\captionsetup{font={footnotesize,sc}, labelsep=newline, justification=centering}
\caption{RA Ratio Inputs for DM and RM}
\label{tab:RA}
 \begin{tabular}{| c || c || c | c || c | c | c |} 
 \hline
Model &
DM &
  \multicolumn{2}{c||}{RM for \eqref{z_cons}} &
   \multicolumn{3}{c|}{RM for \eqref{SOE_min_rob}--\eqref{SOE_nt_rob}}
 \\
 \hline
Parameter &
A &
  $A^{\text{MAX}}$
  & 
 $\Gamma=\Upsilon$
 &
  $A^{\text{MAX}}$
 & 
 $\Upsilon$
    & 
$\Gamma$
 \\
 \hline\
 Input &
 Mean & 
 0.99\% &
  Med & 
 Max  & 
  Min & 
  Max 
\\   
  \hline\hline
UP\_FCR &
0.082 &
\multirow{2}{*}{0.36} &
4 &
\multirow{2}{*}{0.73} &
0.93 &
8.21 
 \\ 
\hhline{-|||-||~|-||~|-|-|}
DN\_FCR &
0.085 &
&
3.8 &
&
1.10
&
16.64 
 \\ 
 \hhline{-|||-||-|-||-|-|-|}
UP\_aFRR &
0.198 &
\multirow{2}{*}{1} &
9.13 &
\multirow{2}{*}{1} &
2.10
&
21.31 
 \\ 
 \hhline{-|||-||~|-||~|-|-|}
DN\_aFRR &
0.218 &
&
9.70 &
&
2.62
&
20.51 
 \\ 
 \hline
\end{tabular}
\end{table}
\vspace{-3mm}
\subsection{Input Parameters} \label{sec:inputs}
The EV driving/parking behaviour was consolidated from the JRC European driving study \cite{JRC-driving}, \cite{JRC-attitude}, and \cite{JRC-projections}. The data was restructured to represent 5-min EV driving/parking behaviour where in each 5-min timestamp each EV can either drive or be parked. Vehicle type and average trip speed were used as inputs to calculate the EV consumption while driving. Starting/finishing trip locations were used to assign instantaneous CP capacity, i.e. to choose the CP type. Three CP types were modeled: low (3.7 kW), medium (7.4 kW) and high (11 kW). Starting/finishing trip times were used to assign whether an EV is driving or is parked. The 5-min timestamps were then summarized into half-hourly periods to match the half-hourly optimization time resolution. As a result, for each EV and each half-hourly period, two input parameters were created: EV driving consumption ($E^{\text{RUN}}_{v,t}$ in eq. \eqref{e_SOE}) and maximum CP power ($P^{\text{CP\_MAX}}_{v,t}$ in eqs. \eqref{f_up} and \eqref{f_dn}).
The data-set for France was used with the total of 581 EVs divided in three types based on vehicle type (battery capacity, OBC size, fleet share): small (20 kWh, 3.7 kW, 30\%), medium (40 kWh, 7.4 kW, 40\%) and large (60 kWh, 11 kW, 30\%). The total fleet capacity is 23.08 MWh, and the total OBC power is 4.26 MW. The EV battery capacity limits are $SOE^{\text{T0}}=0.6$, $SOE^{\text{MAX}}=1$, $SOE^{\text{MIN}}=0.2$.
\vspace{-4mm}
\subsection{Characteristic Days} \label{sec:refdays}
\vspace{-1mm}
The models are tested on four characteristic days:
\begin{itemize}
    \item \textit{Day 1} -- energy price curve with low daily volatility and low FCR capacity price,
    \item \textit{Day 2} -- energy price curve with low daily volatility and high FCR capacity price,
    \item \textit{Day 3} -- energy price curve with high daily volatility and low FCR capacity price,
    \item \textit{Day 4} -- energy price curve with high daily volatility and high FCR capacity price.
\end{itemize}
The prices are taken from the EPEX and from the RTE website for 2018-2019 The aFRR is the same in all four days as its price in France is regulated. 
\begin{comment}
Used prices are displayed at Figure \ref{fig:Prices}.
%
%left bottom right top
\begin{figure}
    \centering
    \includegraphics[trim=20 0 10 10, clip, width=0.8\linewidth]{Inputs.pdf}
    \caption{Used Prices for Model Testing}
    \label{fig:Prices}
\end{figure}
\end{comment}
%
\begin{figure*}[!t]
    \centering
    \includegraphics[trim=115 5 60 5, clip, width=0.8\linewidth]{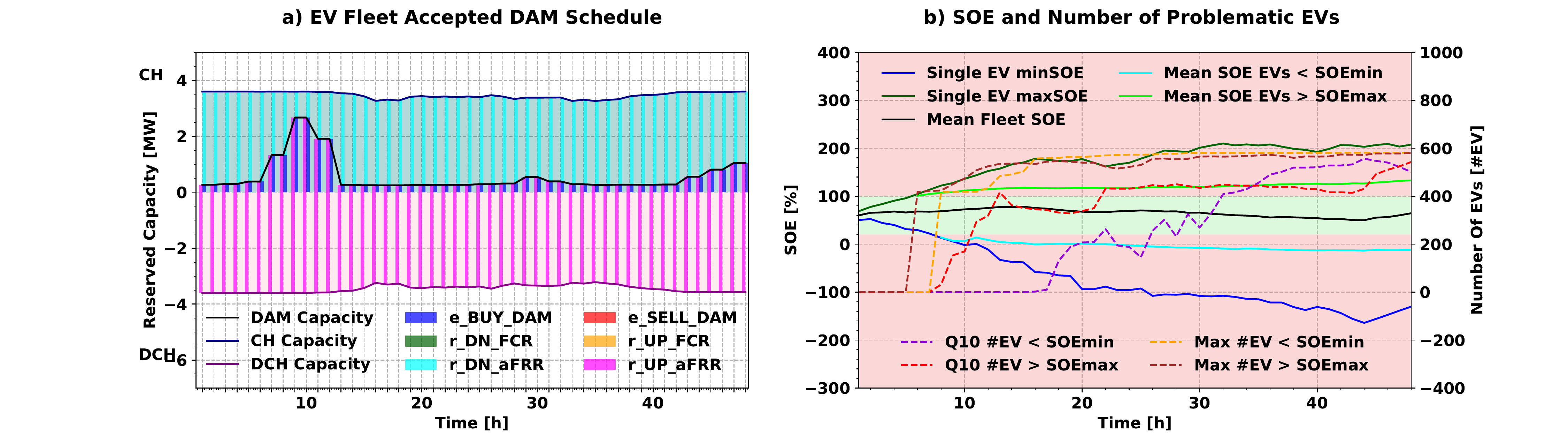}
    \captionsetup{font=footnotesize, labelsep=period, justification   = raggedright}
    \vspace{-1mm}
    \caption{Deterministic Model -- Day 1}
    \label{fig:DC_D1}
\end{figure*}
\begin{figure*}[!t]
    \centering
    \includegraphics[trim=115 5 60 5, clip, width=0.8\linewidth]{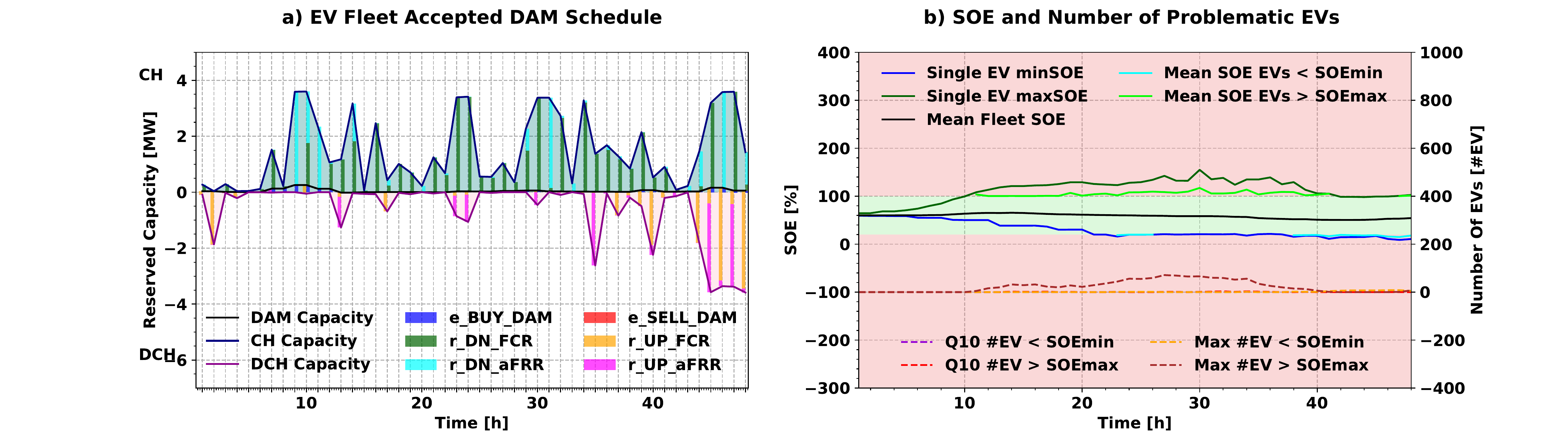}
    \captionsetup{font=footnotesize, labelsep=period, justification   = raggedright}
    \vspace{-1mm}
    \caption{Stochastic Model -- Day 1}
    \label{fig:SC_D1}
\end{figure*}
\begin{figure*}[!t]
    \centering
    \includegraphics[trim=115 5 60 5, clip, width=0.8\linewidth]{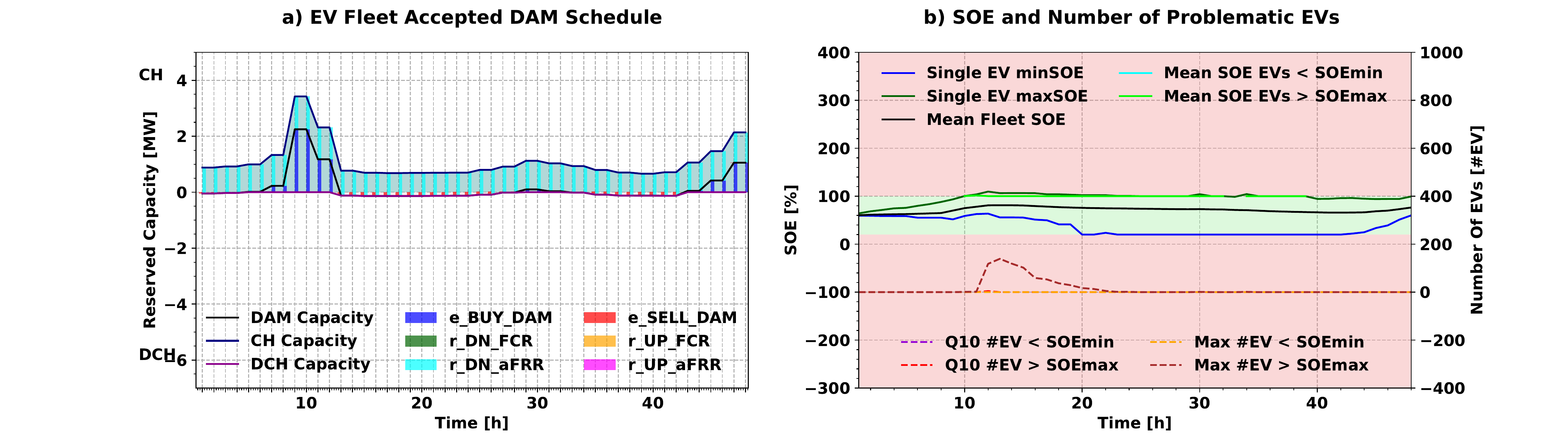}
    \captionsetup{font=footnotesize, labelsep=period, justification   = raggedright}\vspace{-1mm}
    \caption{Robust Model -- Day 1}
    \label{fig:RC_D1}
\vspace{-4mm}
\end{figure*}
\vspace{-2mm}
\section{Results and Discussion} \label{sec:res}
\vspace{-1mm}
The relevant results are shown as time-series in Figures \ref{fig:DC_D1}--\ref{fig:RC_D1} for Day 1 and as aggregates in Table \ref{tab:aggres} for the four characteristic days. They are based on the ex-post simulation using 100 scenarios, each corresponding to a timeseries of the RA ratios. If the reserve is scheduled in the DA, it affects the SOE of a particular EV. If the RA at a certain time-step is such that it steers the SOE to its limits and disables the adherence of the planned DA schedule, the aggregator must redispatch this EV by trading in subsequent intraday/balancing markets to backtrack the SOE to its DA-planned value. The SOE limits cannot be violated as any kind of charging or discharging would stop if an EV reaches the battery limits. However, in this ex-post analysis the violation of the SOE limits is used as an indicator if the EV DA scheduling algorithm incorrectly models the RA uncertainty. The goal of the proposed algorithms is to ensure that even without trading in subsequent markets the RA do not cause infeasible SOE levels.  
\vspace{-3mm}
\subsection{Day-ahead Plans -- Day 1}
\vspace{-1mm}
Figures \ref{fig:DC_D1}--\ref{fig:RC_D1} show the results of the three models. Subfigures to the left show the DA schedules and subfigures to the right show how those plans affect the SOE constraints of the EV fleet. The DM schedules the maximum aFRR (at higher price as compared to the FCR) in both directions plus it schedules a lot of DA charging energy to compensate for the missing energy. %driving energy and for the difference between up and down average activation.
%compensate energy lost due to driving and lost between up and down reserve activations (average values are not the same).
Due to a high amount of the scheduled reserves, a high number of EVs end up with their SOE limits violated in both directions (up to 581) and by a significant margin (up to -164\% and 210\%). In Figure \ref{fig:DC_D1}b) the solid lines denote SOE in \% and the dashed lines represent the number of EVs outside the limits in the worst scenario \textit{Max} and excluding the 10\% of the worst scenarios \textit{Q10}). 
%(solid lines on subfigure b) of Figure \ref{fig:DC_D1} and there is quite a number of EVs in such conditions (dashed lines: the number of EVs outside limits in the worst scenario \textit{Max} and excluding the 10\% of the worst scenarios \textit{Q10}).

The SM adjusts the reserve schedule to its ex-ante scenarios and results in variable reserve schedules where both the FCR and the aFRR are utilized in both directions, as shown in Figure \ref{fig:SC_D1}. The down reserve is scheduled more frequently to avoid charging in the DA energy market. Even though the aFRR reserve is better priced, the FCR is scheduled more since it is less stochastic. Number of EVs beneath the $SOE^{\text{MIN}}$ level is negligible during the entire day, whereas during the daytime certain amount of EVs can surpass $SOE^{\text{MAX}}$  (up to 71 EVs and up to 155\% of SOE). %(subfigure b) of \ref{fig:SC_D1}).

The RM adjust the reserve schedule to find a trade-off between the OF and the worst-case RA. It schedules the reserve rather uniformly through the day and selects only one aFRR in down direction (higher price than FCR), as presented in Figure \ref{fig:RC_D1}. DA energy charging and discharging is utilized to create the optimal working point for each EV, i.e. to maximize reserve provision for the worst-case realization. There is almost no EVs significantly surpassing $SOE^{\text{MAX}}$. The morning peak of such EVs is visible in Figure \ref{fig:RC_D1}b), but their SOE is only slightly above $SOE^{\text{MAX}}$ (110\% as compared to 155\% for the SM and 210\% for the DM). None of the EVs have an SOE beneath $SOE^{\text{MIN}}$ for the RM versus a negligible number for the SM (8 EVs down to -8.83\%) and a significant number for the DM (556 EVs down to -164\%). 
It can be concluded that both the SM and the RM compromise between the revenue and the uncertainty but in their unique way and that EVs rarely surpass their SOE limits. 
\subsection{Summarized Results -- All Days}
The results shown in Table \ref{tab:aggres} are separated in three major segments, providing the statistics on: I) the realized costs, II) the EVs whose SOE is theoretically under $SOE^{\text{MIN}}$, III) the EVs whose SOE is theoretically over $SOE^{\text{MAX}}$. Shaded cells highlight the worst results for a specific reference day, while the best results are in bold text.

As seen in Figures \ref{fig:DC_D1}--\ref{fig:RC_D1}, the DM provides maximum possible reserve at the expense of provision infeasibility. Thus is displays the lowest costs in segment I of Table \ref{tab:aggres}). \textit{Min Cost} for the DM ranges from -2570 to -1860 \EUR{}, while \textit{Max Cost} is in between -290 and -30 \EUR{}. The RM, due to its strict US on the SOE limits but loose US on the OF, provides the highest cost solution. \textit{Min Cost} is always higher than -50 \EUR{} and \textit{Max Cost} than 340. However, the DM would suffer greatly from the penalties related to energy and reserve redispatching in the real time, whereas the RM would be mostly intact by them, i.e. it would be robust to the risks of not being able to provide the scheduled services. The SM is in between those two models, with \textit{Min Cost} ranging from -200 to -600 \EUR{} and \textit{Max Cost} in between 90 and 310 \EUR{}.
\setlength{\tabcolsep}{4.5pt} %
\begin{table*}[!b]
\vspace{-3mm}
\centering
\captionsetup{font={footnotesize,sc}, labelsep=newline, justification=centering}
\caption{Aggregate Results for Two Reference Days Based on 100 EX-post Scenarios Simulation}
\label{tab:aggres}
 \begin{tabular}{| c c c || c c c | c c c | c c c | c c c |} 
 \hline
& \multirow{2}{*}{Observed Results} & & \multicolumn{3}{|c|}{\emph{Day 1}} & \multicolumn{3}{|c|}{\emph{Day 2}} & \multicolumn{3}{|c|}{\emph{Day 3}} & \multicolumn{3}{|c|}{\emph{Day 4}}\\
 \cline{4-15}
 & & & DM  & SM & RM & DM  & SM & RM & DM  & SM & RM & DM  & SM & RM 
\\  
  \hline\hline\hline
\multirow{2}{*}{I)} & Min Cost & [$10^{3}$ \EUR{}] & \textbf{-2.51} & -0.20 & \cellcolor[HTML]{C0C0C0} 0.01 & \textbf{-1.86} &  -0.59 & \cellcolor[HTML]{C0C0C0} -0.05 & \textbf{-2.57} & -0.21 & \cellcolor[HTML]{C0C0C0} 0.04 & \textbf{-1.87} & -0.60 & \cellcolor[HTML]{C0C0C0} -0.03
 \\ 
  \cline{2-15}
& Max Cost & [$10^{3}$ \EUR{}] &  \textbf{-0.07} & 0.28 & \cellcolor[HTML]{C0C0C0} 0.34 & \textbf{-0.29} &  0.09 & \cellcolor[HTML]{C0C0C0} 0.38 & \textbf{-0.03}
 & 0.31 &  \cellcolor[HTML]{C0C0C0} 0.34  & \textbf{-0.29} & 0.12 &  \cellcolor[HTML]{C0C0C0} 0.37
 \\ 
  \hline\hline
\multirow{3}{*}{II)} & Min SOE & [\%] & \cellcolor[HTML]{C0C0C0} -164.11 & 8.83 & \textbf{20.00} & \cellcolor[HTML]{C0C0C0} -37.78 & 4.99 & \textbf{20.00} & \cellcolor[HTML]{C0C0C0} -141.48 & 6.70 & \textbf{20.00} & \cellcolor[HTML]{C0C0C0} -36.81 & 4.76 & \textbf{20.00}
% \\ 
%  \cline{2-15}
%& Q10 Min SOE & [\%] & \cellcolor[HTML]{C0C0C0} -17.56 & 33.29 & 48.95 & 24.03 & 33.62 & 42.85 & \cellcolor[HTML]{C0C0C0} -14.65 & 33.75 & 58.08 & 28.83 & 33.67 & 47.51
 \\ 
    \cline{2-15}
& Max $\Sigma$SOE $<$ & [MWh] & \cellcolor[HTML]{C0C0C0} 25.12 & 0.01 & \textbf{0.00} & \cellcolor[HTML]{C0C0C0} 4.34 & 0.02 & \textbf{0.00} & \cellcolor[HTML]{C0C0C0} 23.97 & 0.01 & \textbf{0.00} & \cellcolor[HTML]{C0C0C0} 3.98 & 0.03 & \textbf{0.00}
% \\ 
%  \cline{2-15}
%& Q10 Max $\Sigma$SOE $<$ & [MWh] & \cellcolor[HTML]{C0C0C0} 7.67 & \textbf{0.00} & \textbf{0.00} & \cellcolor[HTML]{C0C0C0} 0.28 & \textbf{0.00} & \textbf{0.00} & \cellcolor[HTML]{C0C0C0} 6.79 & \textbf{0.00} & \textbf{0.00} & \cellcolor[HTML]{C0C0C0} 0.10 & \textbf{0.00} & \textbf{0.00}
 \\ 
  \cline{2-15}
& Max \#EV $<$ & [\#] & \cellcolor[HTML]{C0C0C0} 581 & 8 & \textbf{0} & \cellcolor[HTML]{C0C0C0} 523 & 21 & \textbf{0} & \cellcolor[HTML]{C0C0C0} 581 & 7 & \textbf{0} & \cellcolor[HTML]{C0C0C0} 520 & 26 & \textbf{0} 
% \\ 
%  \cline{2-15}
%& Q10 Max \#EV $<$ & [\#] & \cellcolor[HTML]{C0C0C0} 556 & 2 & \textbf{0} & \cellcolor[HTML]{C0C0C0} 80 & 3 & \textbf{0} & \cellcolor[HTML]{C0C0C0} 545 & 2 & \textbf{0} & \cellcolor[HTML]{C0C0C0} 60 & 2 & \textbf{0}
\\
 \hline\hline
\multirow{3}{*}{III)} & Max SOE & [\%] & \cellcolor[HTML]{C0C0C0} 209.89 & 154.97 & \textbf{109.60} & \cellcolor[HTML]{C0C0C0} 190.26 & 176.93 & \textbf{106.84} & \cellcolor[HTML]{C0C0C0} 213.27 & 149.01 & \textbf{111.78} & \cellcolor[HTML]{C0C0C0} 188.53 & 181.31 & \textbf{109.37}
% \\ 
%  \cline{2-15}
%& Q10 Max SOE & [\%] & \cellcolor[HTML]{C0C0C0} 134.57 & 76.08 & 93.69 & 97.59 & 72.54 & 87.68 & \cellcolor[HTML]{C0C0C0} 134.85 & 76.65 & 95.95 &  99.72 & 73.17 & 89.16
   \\ 
    \cline{2-15}
&  Max $\Sigma$SOE $>$ & [MWh] & \cellcolor[HTML]{C0C0C0} 16.15 & 0.27 & \textbf{0.15} & \cellcolor[HTML]{C0C0C0} 12.95 & 0.50 & \textbf{0.31} & \cellcolor[HTML]{C0C0C0} 16.15 & 0.36 & \textbf{0.22} & \cellcolor[HTML]{C0C0C0} 12.41 & 0.60 & \textbf{0.10}
%   \\ 
%  \cline{2-15}
%& Q10 Max $\Sigma$SOE $>$ & [MWh] & \cellcolor[HTML]{C0C0C0} 6.70 & \textbf{0.01} & \textbf{0.01} & \cellcolor[HTML]{C0C0C0} 0.31 & 0.01 & \textbf{0.00} & \cellcolor[HTML]{C0C0C0} 6.60 & 0.01 & \textbf{0.00} & \cellcolor[HTML]{C0C0C0} 0.48 & 0.01 & \textbf{0.00}
 \\ 
  \cline{2-15}
& Max \#EV $>$ & [\#] & \cellcolor[HTML]{C0C0C0} 580 & \textbf{71} & 139 & \cellcolor[HTML]{C0C0C0} 572 & \textbf{307} & 420 & \cellcolor[HTML]{C0C0C0} 580 & \textbf{89} & 170 & \cellcolor[HTML]{C0C0C0} 570 & 457 & \textbf{105}
% \\ 
%  \cline{2-15}
%& Q10 Max \#EV $>$ & [\#] & \cellcolor[HTML]{C0C0C0} 543 & \textbf{3} & 4 & \cellcolor[HTML]{C0C0C0} 131 & 4 & \textbf{2} & \cellcolor[HTML]{C0C0C0} 542 & \textbf{3} & 4 & \cellcolor[HTML]{C0C0C0} 156 & 3 & \textbf{2}
 \\ 
 \hline
\end{tabular}
%\vspace{-3mm}
\end{table*}

In segments II) and III) of Table \ref{tab:aggres}, \textit{Min/Max} parameters refer to the worst realization of the observed parameter. 
%, while \textit{Q10 Min/Max} puts aside 10\% of the worst scenarios.
In segment II), \textit{Min SOE} corresponds to the lowest EV SOE realized in the ex-post analysis. The lowest \textit{Min SOE} is always achieved for the DM with -164.11\% in the worst characteristic day (Day 1). The SM yields better results as compared to the DM, but \textit{Min SOE} is still underneath the allowable limits of 20\% (for the worst reference day -- Day 4 it is 4.76\%). The lowest SOE in the RM is just at the lower SOE limit, i.e. 20\%, meaning it does not violate the $SOE^{\text{MIN}}$ limit in any of the observed reference days. %Observing \textit{Q10 Min} it can be seen that DM still can yield SOE bellow limits or close to them, whereas SM and RM achieve high levels of the lowest individual SOE. 
\textit{Max $\Sigma\!<$SOE} indicates the overall energy below $SOE^{\text{MIN}}$ for the entire fleet in one time-step. The DM results in highest energy mismatch, going as high as 25 MWh of unsupplied energy, the SM is in range of several dozens kWh, and the RM is without such energy mismatch.  
%If 10\% of the worst scenarios are removed, the DM can still yield more than 6 MWh of energy mismatch, whereas SM and RM yield zero kWh of mismatch.
\textit{Max \#EV$<$} indicates the number of EVs with SOE falls below $SOE^{\text{MIN}}$ in the worst case. In the DM, all or almost all EVs (520-581) suffer from the SOE lower than $SOE^{\text{MIN}}$ .
%but number of such EVs could be high as 556 even in the case when 10\% of the worst scenarios is removed.
The SM schedule results in up to 26 EVs below $SOE^{\text{MIN}}$, whereas the RM model does not suffer from such issues.

The parameters from segment III) in Table \ref{tab:aggres} are analogous to those from segment II), except they refer to the EVs' upper SOE limits. \textit{Max SOE} is the highest for the DM (up to 213.27\%), closely followed by the SM (up to 181.31\%), and by far the lowest for the RM (up to 111.78\%).
%In the Q10 case, DM can still yield SOE above SOE maximal limits whereas SM and RM are always lower.Also, in Q10 case, the RM achieves the highest SOE closer to 100\% compared to SM. 
High \textit{Max $\Sigma\!>$SOE} values are achieved for the DM (up to 16.15 MWh), whereas the SM and RM are in the range of several hundreds kWh (RM lower for all characteristiv days). 
%In Q10 case, DM comes up to 6.60 MWh of energy above $SOE^{\text{MAX}}$, whereas the SM and RM are negligible (up to 10 kWh). 
Maximum number of EVs above $SOE^{\text{MAX}}$ (\textit{Max \#EV$>$}) is extremely high for the DM (almost all EVs, 570-580), but also relatively high for the SM and RM as well (up to 457 and 420, respectively). For the first three characteristic days the SM even achieves better results than the RM. However, these numbers should be observed in relation to the maximum value of SOE. The high number of EVs above $SOE^{\text{MAX}}$ for the RM means high number of EVs whose SOE is just slightly above $SOE^{\text{MAX}}$. %The RM is modeled in a such way to be as close as possible to $SOE^{\text{MAX}}$, due to model settings slight over SOE is possible. 
%Q10 \textit{Max \#EV} shows that DM can have a lot of EVs outside SOE bounds even if the worst scenarios are excluded. SM and RM have only several EVs outside $SOE^{\text{MAX}}$ in this case.

To conclude this results analysis, from the perspective of the SOE limits violation, the RM provides the best solution closely followed by the SM. The DM is prone to high deviations in actual realization of RA.
%
%
%The worst day for robust model is \textit{reference day 2} where the \textit{"Max $\Sigma$SOE $<$"} and \textit{Max "$\Sigma$SOE $>$"} yield 1.01 and 10.29 MWh, respectively. These results are still better than deterministic but they point out that robust model could yield bad results if realized scenario is outside US limits. On Figure \ref{fig:FCR_prob}, the activation causing those large deviations are left on scatter graph where FCR UP activations are zero and FCR DN large. However, if \textit{"Q10 Max $\Sigma$SOE $<$"} and \textit{"Q10 Max $\Sigma$SOE $>$"} are observed for the same day (they are both zero), it could be concluded that those scenarios are rare and in more than 90\% of the cases the model will work as expected. It is worth to note that stochastic model never yields such high values for those parameters. 

%\textit{Min/Max} SOE corresponds to the minimal/maximal EV SOE realized in ex-post analysis. \textit{Max $\Sigma$SOE} corresponds to the summation of energy missing (for $<$ items) or exceeding (for $>$ items) SOE limits over whole fleet in one timestep. \textit{Max \#EV} corresponds to the number of EVs whose SOE falls behind (for $<$ items) or exceeds (for $>$ items) individual SOE limits..
\vspace{-3mm}
\section{Conclusion} \label{sec:concl}
%
%The analysis in the Section \ref{sec:res} takes into account only case studies where the price for capacity and activation is the same for up and down reserves and where activation price is the same as energy price. Those case studies are considered as base tests for uncertainty models as their reserve outputs tend to be symmetrical during the day where the cross-activations of the two reserve directions mutually mitigate, in part, the downsides of activation uncertainty. In reality, however, those prices can differ and reserve direction yielding higher revenue can be contracted more than the other one. In such case, the SOE misalignment can be even higher and cause more damage to the EV owner or aggregator. The need for stochastic or robust formulation stands out even more is such conditions.

The paper brings a novel approach to modelling uncertainty within scheduling of automatic reserves activation (both aFRR and FCR) in European markets. It proposes stochastic- and robust-based optimization models, where scenarios and uncertainty sets are based on an analysis of real-world historical data. The existing deterministic model fails to properly accommodate the uncertain aspects of the reserve activation. The presented case study clearly demonstrates the advantages of the proposed approaches.

Although claiming high DA benefits, the deterministic model results in extreme individual violations of the EVs' battery SOE limits, ranging from -160\% to above 200\% SOE. Additionally, it results in high fleet-level deviations, from 25 MWh above $SOE^{\text{MAX}}$ to 15 MWh below $SOE^{\text{MIN}}$. These deviations would, in reality, manifest as an inability to provide the scheduled DA energy, activated reserve energy or driving energy. The stochastic and robust formulations decrease the risks related to infeasible reserve activations. Minimum and maximum individual SOE levels achieved in the stochastic model are 4.76\% and 181.31\% of SOE and in the robust model 20\% and 111.78\% SOE. At the fleet level, energy levels below $SOE^{\text{MIN}}$ are negligible, while levels above $SOE^{\text{Max}}$ are in the range of dozens of kWh. Both formulations are technology-agnostic and can be implemented in other algorithms (redispatch measures, other markets, adaptive robust algorithms, etc.) and paired with price, behavior or bid acceptance uncertainties. However, the robust model is more flexible and there are many improvements to be made on top of the the features presented in this paper, e.g. it can be further tailored for specific needs tightening or relaxing specific US parameters. 

%Risk mitigation entails lower potential revenues in financially-good reserve realizations, but at the same time it hedges against higher costs in financially- bad reserve realizations.
%The stochastic model proves to be more efficient in certain aspects compared to robust. However, stochastic formulation is close to its limits and it can be only slightly improved (e.g. the scenarios could be generated and reduced in a more concise manner to capture the uncertainty). Also, it is hard to couple those scenarios with other uncertainties which could be part of model as wells such as prices (different energy markets, reserves, activated energy etc.), acceptance of certain reserve bids, EV behavior (arrival, departure, SOE etc.), EV failures etc. The robust model, on the other hand, could be tailored to specific conditions (by changing parameters of the US) and it could be easily expanded to include other uncertainties. The US could be tighten for some needs, whereas relax for the other. For example, the US of OF could be decoupled from the US of the SOE constraints where we can take tighter formulation for constraints and more relaxed version for OF. Also, the second stage variables such as sequential markets or redispatch among EVs could be added in the form of adaptive robust model. To conclude, with robust model there are still many improvements to be made on top of the those presented in this paper.      

\vspace{12pt}


\vspace{-3mm}
\begin{thebibliography}{00}
\bibitem{IEA2019} International Energy Agency, ``Global EV Outlook 2019,'' Technical Report, 2019.
\bibitem{Irena} International Renewable Energy Agency, Innovation landscape brief: Electric-vehicle smart charging, Technical Report, 2019.

\bibitem{JRC-driving}
G. Pasaoglu, D. Fiorello, A. Martino, G. Scarcella, A. Alemanno, A. Zubaryeva, an C. Thiel, ``Driving and parking patterns of European car drivers - a mobility survey,'' European Commission Report, 2012.

\bibitem{ENTSOE} ENTSO-E, ``ENTSO-E Balancing Report,'' 2020.

\bibitem{ACER} ACER, ``ACER Market Monitoring Report 2019 – Electricity Wholesale Markets Volume,'' ACER Market Monitoring Report 2018 –- Electricity Wholesale Markets Volume, 2019.

\bibitem{Figgener} J. Figgener \emph{et al.}, ``The development of stationary battery storage systems in Germany –- A market review,'' \emph{J. Energy Storage}, vol. 29, p. 101153, Jun. 2020.
\bibitem{Newtech} Ramboll, ``Ancillary Services From New Technologies,'' Technical Report, 2019.

\bibitem{PhDbat} J. Engels, ``Integration of Flexibility from Battery Storage in the Electricity Market,'' PhD Thesis, 2020.
%\bibitem{EDSO2018} European Distribution System Operators for Smart Grids, “Smart charging: integrating a large widespread of electric cars in electricity distribution grids,” 2018.
%\bibitem{Eurelectric2017} Eurelectric, “Smart Charging – Key to unlocking Electro-mobility’s potential,” Brussels, 2017.
%\bibitem{Kolt} N. E. Koltsaklis, A. S. Dagoumas, and I. P. Panapakidis, “Impact of the penetration of renewables on flexibility needs,” Energy Policy, vol. 109, pp. 360–369, Oct. 2017.
%\bibitem{NREL} E. Ela, M. Milligan, and B. Kirby, “Operating Reserves and Variable Generation,” Cole Boulevard Golden, 2011.
%\bibitem{MM} M. MacDonald, “Impact Assessment on European Electricity Balancing Market,” Brighton, 2013.
%\bibitem{EE} “Frequency Containment Reserve (FCR).” [Online]. Available: https://www.emissions-euets.com/internal-electricity-market-glossary/793-frequency-containment-reserve. [Accessed: 30-Nov-2019].
%\bibitem{NPV} O. Borne, M. Petit, and Y. Perez, “Net-Present-Value Analysis for Bidirectional EV Chargers Providing Frequency Containment Reserve,” in 2018 15th International Conference on the European Energy Market (EEM), 2018, pp. 1–6.
%\bibitem{Soares} T. Soares, T. Sousa, P. B. Andersen, and P. Pinson, “Optimal Offering Strategy of an EV Aggregator in the Frequency-Controlled Normal Operation Reserve Market,” in 2018 15th International Conference on the European Energy Market (EEM), 2018, pp. 1–6.
%\bibitem{Hashemi} S. Hashemi, N. B. Arias, P. Bach Andersen, B. Christensen, and C. Traholt, “Frequency Regulation Provision Using Cross-Brand Bidirectional V2G-Enabled Electric Vehicles,” in 2018 IEEE International Conference on Smart Energy Grid Engineering (SEGE), 2018, pp. 249–254.
%\bibitem{Herre} L. Herre, J. Dalton, and L. Soder, “Optimal Day-Ahead Energy and Reserve Bidding Strategy of a Risk-Averse Electric Vehicle Aggregator in the Nordic Market,” in 2019 IEEE Milan PowerTech, 2019, pp. 1–6.
%\bibitem{JRC}
%“EU Science Hub - European Commission.” [Online]. Available: https://ec.europa.eu/jrc/en. [Accessed: 30-Nov-2018].

\bibitem{Alipour} M. Alipour \emph{et al.}, ``Stochastic scheduling of aggregators of plug-in electric vehicles for participation in energy and ancillary service markets,'' \emph{Energy}, vol. 118, pp. 1168–-1179, Jan. 2017.

\bibitem{Shafie} M. Shafie-Khah, M. P. Moghaddam, M. K. Sheikh-El-Eslami, and J. P. S. Catalao, ``Optimised performance of a plug-in electric vehicle aggregator in energy and reserve markets,'' \emph{Energy Convers. Manag.}, vol. 97, pp. 393–-408, Jun. 2015.

\bibitem{Momber} I. Momber \emph{et al.}, ``Risk averse scheduling by a PEV aggregator under uncertainty,'' \emph{IEEE Trans. Power Syst.}, vol. 30, no. 2, pp. 882–-891, Mar. 2015.

\bibitem{Vag} S. I. Vagropoulos and A. G. Bakirtzis, ``Optimal bidding strategy for electric vehicle aggregators in electricity markets,'' \emph{IEEE Trans. Power Syst.}, vol. 28, no. 4, pp. 4031–-4041, 2013.

\bibitem{Sanchez} P. Sanchez-Martin, S. Lumbreras, and A. Alberdi-Alen, ``Stochastic programming applied to ev charging points for energy and reserve service markets,'' \emph{IEEE Trans. Power Syst.}, vol. 31, no. 1, pp. 198--205, Jan. 2016.

\bibitem{Bessa}  R. J. Bessa and M. A. Matos, ``Optimization models for EV aggregator participation in a manual reserve market,'' \emph{IEEE Trans. Power Syst.}, vol. 28, no. 3, pp. 3085--3095, 2013.

\bibitem{Goebel}  C. Goebel and H. A. Jacobsen, ``Aggregator-Controlled EV Charging in Pay-as-Bid Reserve Markets with Strict Delivery Constraints,'' \emph{IEEE Trans. Power Syst.}, vol. 31, no. 6, pp. 4447–-4461, Nov. 2016.

\bibitem{Hasanpor} P. Hasanpor Divshali and C. Evens, ``Optimum Operation of Battery Storage System in Frequency Containment Reserves Markets,'' \emph{IEEE Trans. Smart Grid}, vol. 11, no. 6, pp. 4906–-4915, Nov. 2020.

\bibitem{aFRR} M. Merten \emph{et al.}, ``Bidding strategy for battery storage systems in the secondary control reserve market,'' \emph{Appl. Energy}, vol. 268, p. 114951, 2020.

\bibitem{Energies} K. Pand\v{z}i\'c, I. Pavi\'c, I. Andro\v{c}ec, and H. Pand\v{z}i\'c, ``Optimal Battery Storage Participation in European Energy and Reserves Markets,'' \emph{Energies}, 2020.

\bibitem{Liu}  G. Liu, Y. Xu, and K. Tomsovic, ``Bidding strategy for microgrid in day-ahead market based on hybrid stochastic/robust optimization,'' \emph{IEEE Trans. Smart Grid}, vol. 7, no. 1, pp. 227–-237, Jan. 2016.

%\bibitem{Kazemi} M. Kazemi, H. Zareipour, N. Amjady, W. D. Rosehart, and M. Ehsan, “Operation Scheduling of Battery Storage Systems in Joint Energy and Ancillary Services Markets,” IEEE Trans. Sustain. Energy, vol. 8, no. 4, pp. 1726–1735, Oct. 2017.

\bibitem{Kazemi} M. Kazemi \emph{et al.}, ``Operation Scheduling of Battery Storage Systems in Joint Energy and Ancillary Services Markets,'' \emph{IEEE Trans. Sust. Energy}, vol. 8, no. 4, pp. 1726--1735, Oct. 2017.

\bibitem{Sarker} M. R. Sarker, Y. Dvorkin, and M. A. Ortega-Vazquez, ``Optimal participation of an electric vehicle aggregator in day-ahead energy and reserve markets,'' \emph{IEEE Trans. Power Syst.}, vol. 31, no. 5, pp. 3506--3515, Sept. 2016.

\bibitem{Ivan} I. Pavi\'c, H.  Pand\v{z}i\'c, T. Capuder, ``Electric Vehicles as Frequency Containment Reserve Providers,'' in \emph{Proceedings of IEEE International Energy Conference (ENERGYCON)}, Gammarth, Tunisia, Oct. 2020.

%\bibitem{Sarker} M. R. Sarker, Y. Dvorkin, and M. A. Ortega-Vazquez, “Optimal participation of an electric vehicle aggregator in day-ahead energy and reserve markets,” IEEE Trans. Power Syst., vol. 31, no. 5, pp. 3506–3515, Sep. 2016.


%\bibitem{Han} B. Han, S. Lu, F. Xue, and L. Jiang, “Day-ahead electric vehicle aggregator bidding strategy using stochastic programming in an uncertain reserve market,” IET Gener. Transm. Distrib., vol. 13, no. 12, pp. 2517–2525, Jun. 2019.

\bibitem{Han} B. Han \emph{et al.}, ``Day-ahead electric vehicle aggregator bidding strategy using stochastic programming in an uncertain reserve market,'' \emph{IET Gener. Transm. Distrib.}, vol. 13, no. 12, pp. 2517--2525, Jun. 2019.

\bibitem{Ortega} M. A. Ortega-Vazquez, ``Optimal scheduling of electric vehicle charging and vehicle-to-grid services at household level including battery degradation and price uncertainty,'' \emph{IET Gener. Transm. Distrib.}, vol. 8, no. 6, pp. 1007-–1016, June 2014.

%\bibitem{Yan}  Y. Cui, Z. Hu, and H. Luo, “Optimal Day-Ahead Charging and Frequency Reserve Scheduling of Electric Vehicles Considering the Regulation Signal Uncertainty,” IEEE Trans. Ind. Appl., vol. 56, no. 5, pp. 5824–5835, Feb. 2020.

\bibitem{RTE} RTE, ``RTE Customer’s area - Volumes and prices.'' [Online]. Available: https://clients.rte-france.com [Accessed: 22-Jan-2019].

\bibitem{JRC-attitude}
C. Thiel, A. Alemanno, G. Scarcella, A. Zubareyeva, and G. Pasaoglu, ``Attitude of European car drivers towards electric vehicles: a survey,'' JRC Report, 2012.

\bibitem{JRC-projections}
G. Pasaoglu \emph{et al.}, ``Projections for EV Load Profiles in Europe Based on Travel Survey Data Contact information,'' JRC Report, 2013.

\bibitem{Morales}
J. M. Morales \emph{et al.}, Integrating Renewables in Electricity Markets, vol. 205. Boston, MA: Springer US, 2014.
\end{thebibliography}
\end{document}